\documentclass[iop,apj,tighten,numberedappendix,twocolappendix]{emulateapj}
\usepackage{apjfonts}
\usepackage{psfig}

\usepackage{amsmath}
\usepackage[normalem]{ulem}
\usepackage{graphicx}
	\DeclareGraphicsExtensions{.eps, .jpg}
\usepackage{color}
	\definecolor{gray}{rgb}{0.6,0.6,0.6}
	\definecolor{green}{rgb}{0,0.6,0}
\newcommand{\Msun}{M_{\odot}}
\newcommand{\Mdot}{\dot{M}}
\newcommand{\MdotEdd}{\dot{M}_{\rm Edd}}
\newcommand{\LEdd}{L_{\rm Edd}}

\input epsf.sty

\slugcomment{Accepted for publication in \textit{The Astrophysical Journal}}

\shorttitle{Winds in X-ray binaries}
\shortauthors{You et al.}

\begin{document}

\title{
Testing wind as an explanation for the spin problem in the continuum-fitting method
}

\author{
Bei You\altaffilmark{1},
Odele Straub \altaffilmark{2,3},
Bo\.zena Czerny\altaffilmark{1,4},
Ma\l gosia Sobolewska\altaffilmark{1,5},
Agata R\' o\. za\' nska\altaffilmark{1},
Michal Bursa \altaffilmark{6},
Michal Dov\v ciak \altaffilmark{6}
}
\altaffiltext{1}{
N. Copernicus Astronomical Center, Bartycka 18, 00-716 Warsaw, Poland;
youbeiyb@gmail.com, bcz@camk.edu.pl, malgosia@camk.edu.pl, agata@camk.edu.pl
} 
\altaffiltext{2}{
Astrophysics, Department of Physics, University of Oxford, Keble Road, Oxford OX1 3RH, UK; odelest@protonmail.com
} 
\altaffiltext{3}{
LUTH, CNRS UMR 8102, Observatoire de Paris, Universite Paris Diderot, 92190 Meudon, France
} 
\altaffiltext{4}{
Center for Theoretical Physics, Al. Lotnik\'ow 32/46, 02-668 Warsaw, Poland
} 
\altaffiltext{5}{
Harvard-Smithsonian Center for Astrophysics, 60 Garden Street, Cambridge, MA 02138, USA
} 
\altaffiltext{6}{
Astronomical Institute, Academy of Sciences, Bo\v cn\'i II 1401, 14131 Prague, Czech Republic; bursa@astro.cas.cz, dovciak@asu.cas.cz
} 

\begin{abstract}
The continuum-fitting method is one of the two most advanced methods of determining the black hole spin in accreting X-ray binary systems. There are, however, still some unresolved issues with the underlying disk models. One of them manifests as an apparent decrease in spin for increasing source luminosity.
Here, we perform a few simple tests to establish whether outflows from the disk close to the inner radius can address this problem.
We employ four different parametric models to describe the wind and compare these to the apparent decrease in spin with luminosity measured in the sources LMC~X-3 and GRS~1915+105. 
Wind models in which parameters do not explicitly depend on the accretion rate cannot reproduce the spin measurements. Models with mass accretion rate dependent outflows, however, have spectra that emulate the observed ones.
The assumption of a wind thus effectively removes the artifact of spin decrease. This solution is not unique; the same conclusion can be obtained with a truncated inner disk model. To distinguish among valid models, high resolution X-ray data and a realistic description of the Comptonization in the wind will be needed. 
\end{abstract}


\section{Introduction}
\label{sec:introduction}
The determination of the black hole spin is one of the most intriguing issues in astrophysics. Knowledge of the spin sheds light on the evolution of X-ray binary systems (XRBs) and active galactic nuclei (AGN) \citep{2008MNRAS.385.1621K, 2008MNRAS.383.1079V, 2009ApJ...697L.141W, 2010ApJ...725..388C}, and it allows to assess the role of the black hole angular momentum in the formation of relativistic jets \citep{1977MNRAS.179..433B, 2001MNRAS.328L..27W, 2002ApJ...570L..69M, 2011ApJ...735...50W, 2012ApJ...753...24L, 2012MNRAS.419L..69N, 2015ApJS..221....3G}. There are several ways to determine the spin of a black hole. The two most reliable methods to date are fitting the broad Fe K$_{\alpha}$ line from reflected emission \citep{2006ApJ...652.1028B, 2008ApJ...679L.113M, 2015MNRAS.447..517L} and fitting the continuum emission \citep{1997ApJ...482L.155Z, 2014ApJ...790...29G,2014ApJ...793L..29S}. Further ways to estimate the black hole spin include X-ray timing, i.e. the modeling of high-frequency quasi-periodic oscillations (HFQPOs) \citep[][]{1999ApJ...524L..63S, 2014MNRAS.437.2554M, 2014Natur.513...74P} and X-ray polarimetry of the inner disk region \citep{1980ApJ...235..224C, 2008MNRAS.391...32D, 2009ApJ...691..847L}. A detailed discussion of these four methods can be found in the review by \citet{2009astro2010S.197M}. In this paper we concentrate on the problems connected to the continuum-fitting method.

Spin measurements for a number of (mostly) Galactic XRB sources based on the continuum-fitting method are available in the literature, for instance for GRS~1915+105 \citep[e.g.][]{2006ApJ...652..518M, 2006MNRAS.373.1004M} and LMC~X-3 \citep[e.g.][]{2006ApJ...647..525D, 2010ApJ...718L.117S}, two sources that attain very high luminosities. The method was also successfully used in the case of the Weak Line Quasar SDSS J094533.99+100950.1 \citep{2011MNRAS.415.2942C}, the radio-loud narrow-line Seyfert 1 galaxy RX J1633+4718 \citep{2010ApJ...723..508Y}, and several other intermediate redshift quasars \citep{2015MNRAS.446.3427C}. 

Despite the overall success of the continuum-fitting method, one fundamental problem remains unsolved, namely the apparent dependence of the spin on the source luminosity. During an outburst, XRBs go through the high/soft state where their spectra are dominated by emission that peaks at soft X-ray energies. This peak can be modeled by a multi-color blackbody disk. Thermal spectra with a disk flux contribution above $\geq 75\%$ to the total flux are suitable for the analysis \citep{2006ARA&A..44...49R}. The continuum-fitting method compares the observed shape of the thermal continuum spectrum to an accretion disk model spectrum. A sequence of such thermal state spectra have been modeled, e.g. for GRS~1915+105 \citep{2006ApJ...652..518M}. There, the fits resulted in a black hole spin that was significantly decreasing with the modeled disk luminosity above  $L \gtrsim 0.3 \LEdd$, where $\LEdd = 1.26 \times 10^{38} \ (M/\Msun)$ erg/s.

A plausible explanation is that the assumptions underlying the disk model in \citet{2006ApJ...652..518M} break down at higher Eddington ratios. The standard accretion model describes an optically thick and geometrically razor thin ($H/R \ll 1$) gas disk that reaches all the way down to the innermost stable circular orbit (ISCO) and efficiently emits at each annulus \citep{1973A&A....24..337S, 1973blho.conf..343N}. With increasing mass accretion rate the standard disk is unable to cool efficiently. As the disk overheats, it expands vertically to scale heights that are not consistent with the initial assumptions anymore and a significant fraction of energy is advected inward. The disk structure is then better described by the slim disk model \citep{1988ApJ...332..646A, 2011A&A...527A..17S}, which has a radiative efficiency that decreases with increasing mass accretion rate. The corresponding analysis has been done, e.g. for LMC~X-3 \citep{2011A&A...533A..67S}. The estimated black hole spin, however, {\it still} decreases with luminosity also for the advective slim disk. 

There are two reasons which could explain this apparent decrease in the spin. The first one is the description of the disk atmosphere. Radiative transfer in the disk atmosphere is usually incorporated in accretion disk spectral models via the TLUSTY code \citep{1995ApJ...439..875H} which locally computes absorption opacities of various chemical elements in plane-parallel layers. In the X-ray spectral fitting software XSPEC \citep{1996ASPC..101...17A} a well known and widely used thin disk model with full radiative transfer is {\sc bhspec} \citep{2006ApJS..164..530D}, while slim disks with full radiative transfer are available as {\sc slimbh}\footnote{\url{http://astro.cas.cz/slimbh}} \citep{2011A&A...527A..17S, 2011A&A...533A..67S}. However, the physics of the disk atmosphere is very complicated and largely unknown, with magnetic fields, turbulent stress, and dissipation taking place \citep[see e.g][]{2011ApJ...733..110B,2013ApJ...770...55T}. This manifests in the fact that both {\sc bhspec} and {\sc slimbh} seem to overestimate the amount of photons at the spectral peak \citep[see Fig.~4 in][]{2013A&A...553A..61S}. The second reason is mass-loss due to winds/outflows from the disk surface. In general, outflows are thought to become significant when the mass accretion rate through the disk increases as discussed in a number of papers \citep{2003MNRAS.345..657K, 2012MNRAS.426..656S, 2012ApJ...761..130Y, 2014ApJ...783...51C, 2014MNRAS.438.3024L, 2015ApJ...804..101Y,2016ApJ...817...71C}. Outflowing winds have recently been detected in several XRBs and AGNs via blueshifted X-ray absorption lines which are observed in high resolution spectra, for instance in GRS~1915+105 \citep[e.g.][]{2009Natur.458..481N}, GRO J1655$-$40 \citep[e.g.][]{2006Natur.441..953M, 2009ApJ...701..865K}, IGR J17091-3624 \citep[e.g.][]{2012ApJ...746L..20K}, NGC 3783 \citep[e.g.][]{2002ApJ...574..643K}, and PG 1211+143 \citep[e.g.][]{2003MNRAS.345..705P}.

In the present paper we perform simple estimates to check whether a wind scenario is likely to be responsible for the observed apparent spin decrease over luminosity. Given the fact that spin measurements from both thin and slim disks show the same type of deviation we use the mathematically simpler standard \citet{1973blho.conf..343N} model to describe the underlying accretion disk. We then modify the mass accretion rate to incorporate mass loss due to an unspecified wind and calculate the energy spectra of such flows. In Sec.~\ref{sec:estim} and \ref{sec:flow} we present an estimate of the energy that is driven away by wind and possible disk models with wind. Sec.~\ref{sec:comp} shows Comptonization of the disk spectrum by the wind. In the extended discussion in Sec.~\ref{sec:discussion} we test our models against observational data of the two most explored sources LMC~X-3 and GRS~1915+105. We give our conclusions in Sec.~\ref{sec:conl}.

\section{Analytical estimates}
\label{sec:estim}
\begin{figure}
\centering
\includegraphics[width=0.45\textwidth]{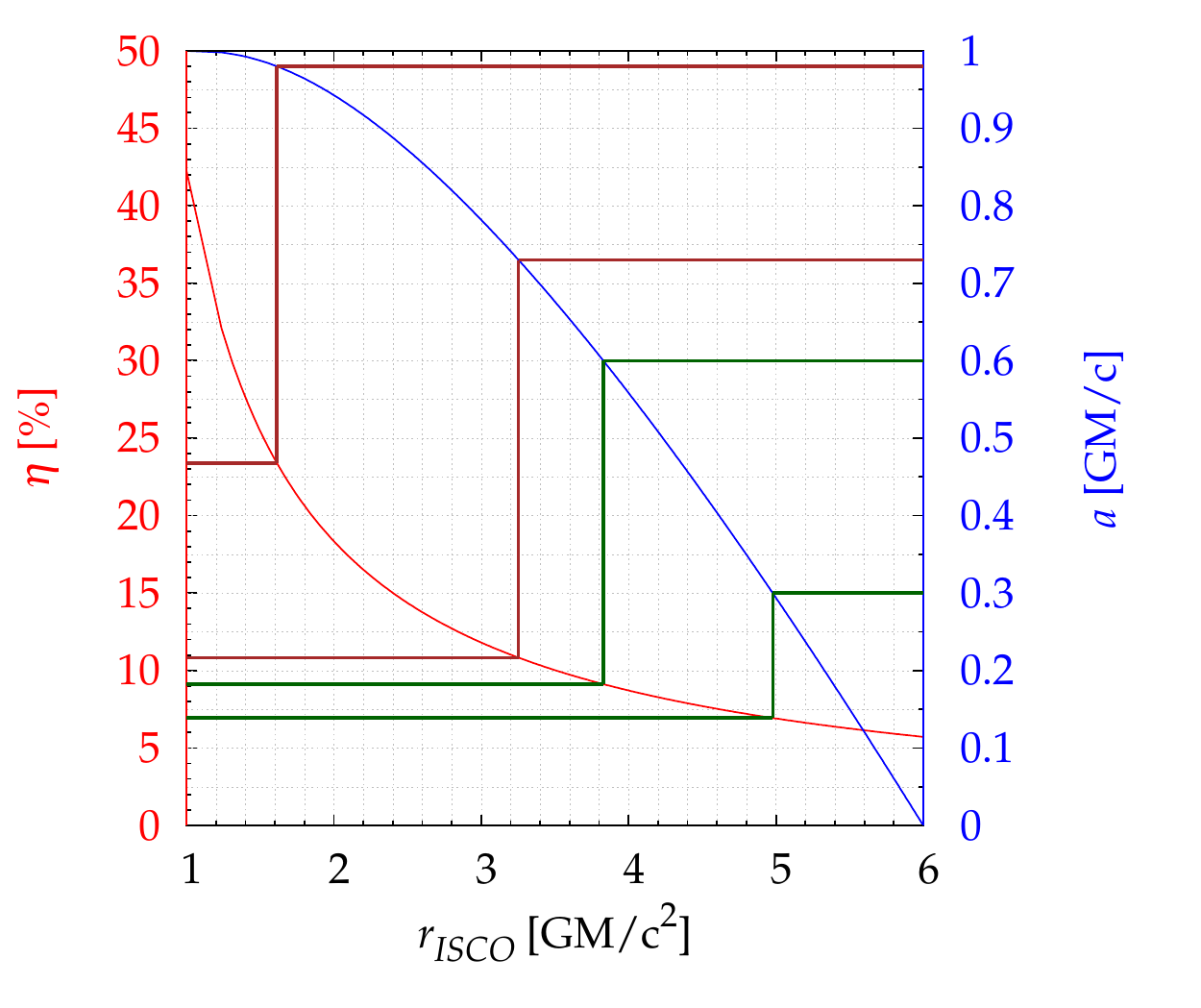}
\caption{
Dependences of the theoretical mass to radiation conversion efficiency, $\eta$, of a relativistic standard disk on the black hole spin, $a$, and the inner disk radius, $r_{\rm ISCO}$. The efficiency that corresponds to the measured spins in GRS~1915+105 and LMC~X-3 is highlighted in brown and dark green, respectively. 
}
\label{fig:efficiency}
\end{figure}

The apparent decrease of the black hole spin with increasing source luminosity reported in GRS~1915+105 and LMC~X-3 is significant in both cases \citep{2006ApJ...652..518M, 2010ApJ...718L.117S, 2011A&A...533A..67S}. For this reason the disk is usually assumed to be reliably optically thick and geometrically thin only for modeled disk luminosities in the range $L \approx 0.05 - 0.3 \ \LEdd$. In the relativistic standard model \citep{1973blho.conf..343N}, the apparent decline of spin with luminosity can be directly translated into a decreasing accretion efficiency, $\eta$. We now postulate that (i) this decrement in spin roots in the existence of wind and that (ii) the amount of energy driven away by this wind corresponds to the drop in accretion efficiency. The efficiency to convert mass into radiation is given by
\begin{equation}\label{efficiency}
	\eta = 1 - U_t(r_{\rm ISCO}),
\end{equation}
where $U_t(r_{\rm ISCO}) = ( r_{\rm ISCO}^2 - 2 r_{\rm ISCO} + a r_{\rm ISCO}^{1/2} ) / ( r_{\rm ISCO}^4 - 3 r_{\rm ISCO}^3 + 2 a r_{\rm ISCO}^{5/2} )^{1/2}$ is the co-variant time component of the four-velocity of the gas. For convenience, we show the theoretical radiation efficiency for a relativistic thin disk in Fig.~\ref{fig:efficiency}. A more realistic modeling of how efficiency depends on spin one can be found in \citet{2005ApJS..157..335L}. Wind then has to carry away the fraction $\Delta \eta $ of the total accretion energy
\begin{equation}
	\Delta \eta = \eta_0 - \eta,
\end{equation}
where $\eta_0$ is the asymptotic flow efficiency at low luminosity. 

Recent optical data lead to an improvement of the dynamical parameters of LMC~X-3 \citep{2014ApJ...794..154O}. The black hole mass, $M \approx 6.98 \Msun$ and the inclination angle, $i \approx 69.2^{\circ}$, result in a black hole spin, $a \approx 0.25$ \citep[][]{2014ApJ...793L..29S}. In the framework of our forthcoming paper II (Straub et al., in preparation) we estimate the black hole spin in LMC~X-3 based on the improved parameters using the slim disk model, following the methodology of \cite{2011A&A...533A..67S}. We find for $L \leq 0.3 L_{\rm Edd}$ a mean spin value $a = 0.2$ that trails off to $a = 0$ very quickly beyond $L \approx 0.3 \LEdd$ which is consistent with previous work. Given the flat theoretical profile of $\eta$ at small spins, the new values imply $\Delta \eta = 6.46 - 5.72 = 0.74\%$. 

Recent measurements of the GRS~1915+105 parameters yield a black hole mass of $M \approx 12.4 \Msun$, an inclination of $i \approx 60^{\circ}$, and a source distance of about 8.6 kpc \citep{2014ApJ...796....2R}. These values entail a spin of $a \approx 0.95$ \citep{2015ApJ...800...17F} which falls off to $a \approx 0.86$ (assuming $\alpha=0.1$, Straub et al., in preparation). This translates to an asymptotic efficiency $\eta_0 = 19.01\%$ that decreases to $\eta = 13.92\%$ resulting in $\Delta \eta \simeq 5.09\%$. 

We adopt throughout the paper the above quoted dynamical parameters for LMC~X-3 and GRS~1915+105 and the respective mean spin values which we found using the slim disk model.

\section{Simple wind models}
\label{sec:flow}
In this section we introduce a simple parametric description of the wind. First, we concentrate on the effect the wind has on the intrinsic shape of the thermal emission from an accretion disk. We describe the wind as a radially dependent quantity that decreases the accretion rate towards the black hole. In the following calculations we adopt the code by \citet{2011MNRAS.415.2942C} which is based on \citet{1973blho.conf..343N} and neglects self-irradiation of the disk as well as Comptonization in the disk atmosphere. Moreover, we assume that the disk thickness is small so that the photons are emitted from the equatorial plane \citep{2009ApJS..183..171S}. Therefore, the local disk emissivity is independent of any assumptions about the disk's vertical structure, including viscosity. We assume zero torque at the ISCO. 

The wind from the accretion disk surface that is required to cause the apparent change in the measured spin must have the appropriate radial distribution. Since the observed X-ray emission comes predominantly from the innermost tens of gravitational radii, the type of wind that can solve the spin drop problem must be localized there.

There are a number of ways to describe the mass loss due to wind. We define $\Mdot_{\rm in}$ and $\Mdot_{\rm out}$ as the accretion rates at the inner and outer disk radius, $r_{\rm in}$ and $r_{\rm out}$, respectively. $\Mdot(r)$ and $\Mdot_{\rm wind}(r)$ are the inflow at a given radius, $r$, and total mass loss beyond this radius, respectively. We consider several wind models and calculate their spectra for a range of accretion rates $\Mdot_{\rm out}$ and fixed spin parameter $a$. Then, we fit the emergent spectra using the relativistic standard model \citep{1973blho.conf..343N} {\it without} any wind, leaving accretion rate $\Mdot_{\rm out}^{\prime}$ and spin $a^{\prime}$ as free parameters. This procedure allows to verify whether the measured spin is lower than the input spin in the disk-wind model as it is in the fits to observational data, i.e. if $a^{\prime} < a$. 

\begin{figure}[ht!]
\centering
\includegraphics[width=0.45\textwidth]{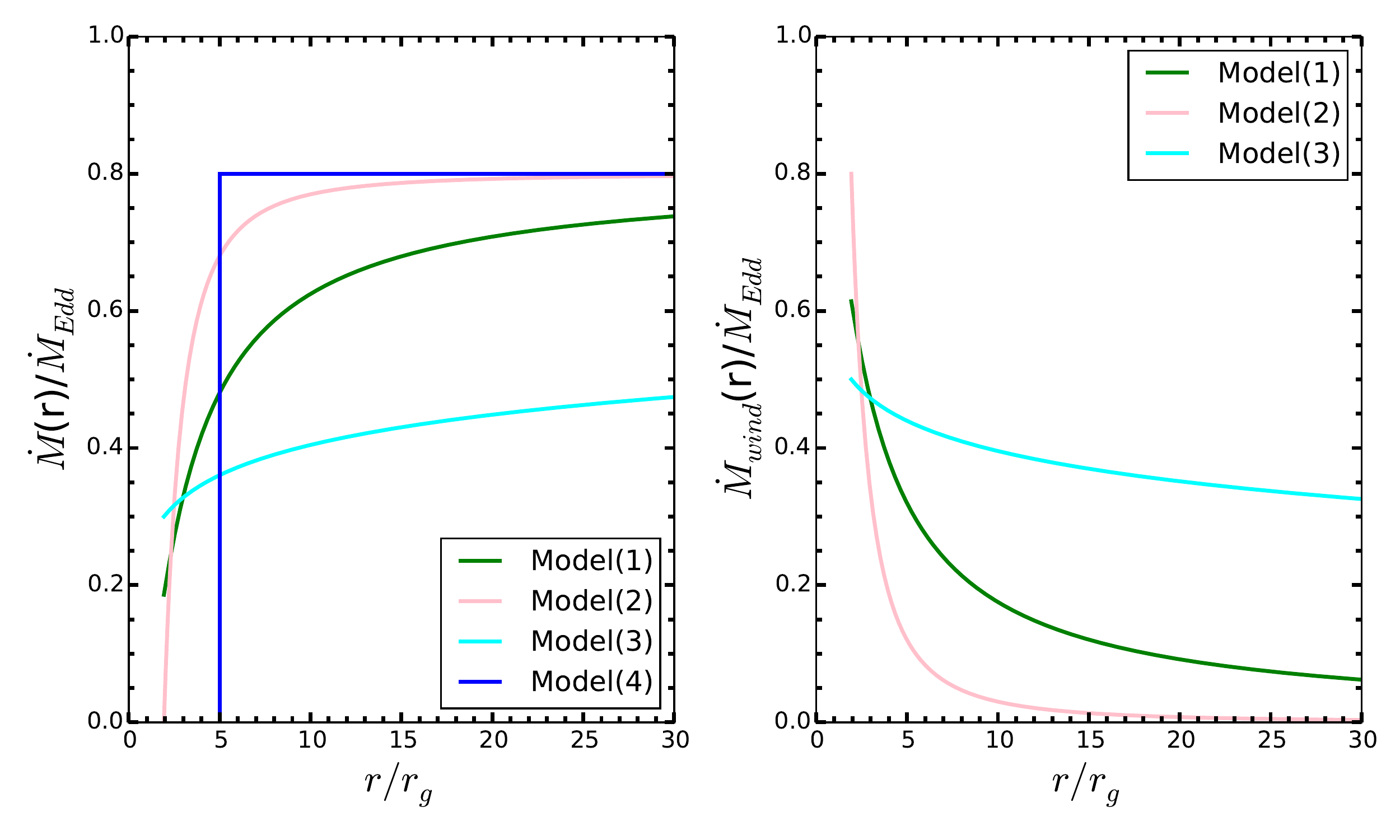}
\caption{
{\it Left}: Radial profiles of the mass accretion rate for the four postulated outflow models. {\it Right}: Radial profiles of the mass outflow rate. In Models (1)-(3) the radial distribution of the mass accretion rate, $\Mdot$, is given by an exponential function, a power law, and a differential function. In Model~(4) $\Mdot(r)$ is constant beyond the truncation radius. We globally assume $\Mdot_{\rm out}/\MdotEdd = 0.8$, $a = 0.95$, and $r_{\rm in} = r_{\rm ISCO}$ for Models (1)-(3) and $r_{\rm in} = 5 \ r_{\rm g}$ for Model~(4). In addition, we employed $A = 1.24$, $B = 0.15$, and $\beta = -2.0$ (see Sec.~\ref{sec:flow}).
}
\label{fig:outflow}
\end{figure}

\begin{figure}[ht!]
\includegraphics[width=0.45\textwidth]{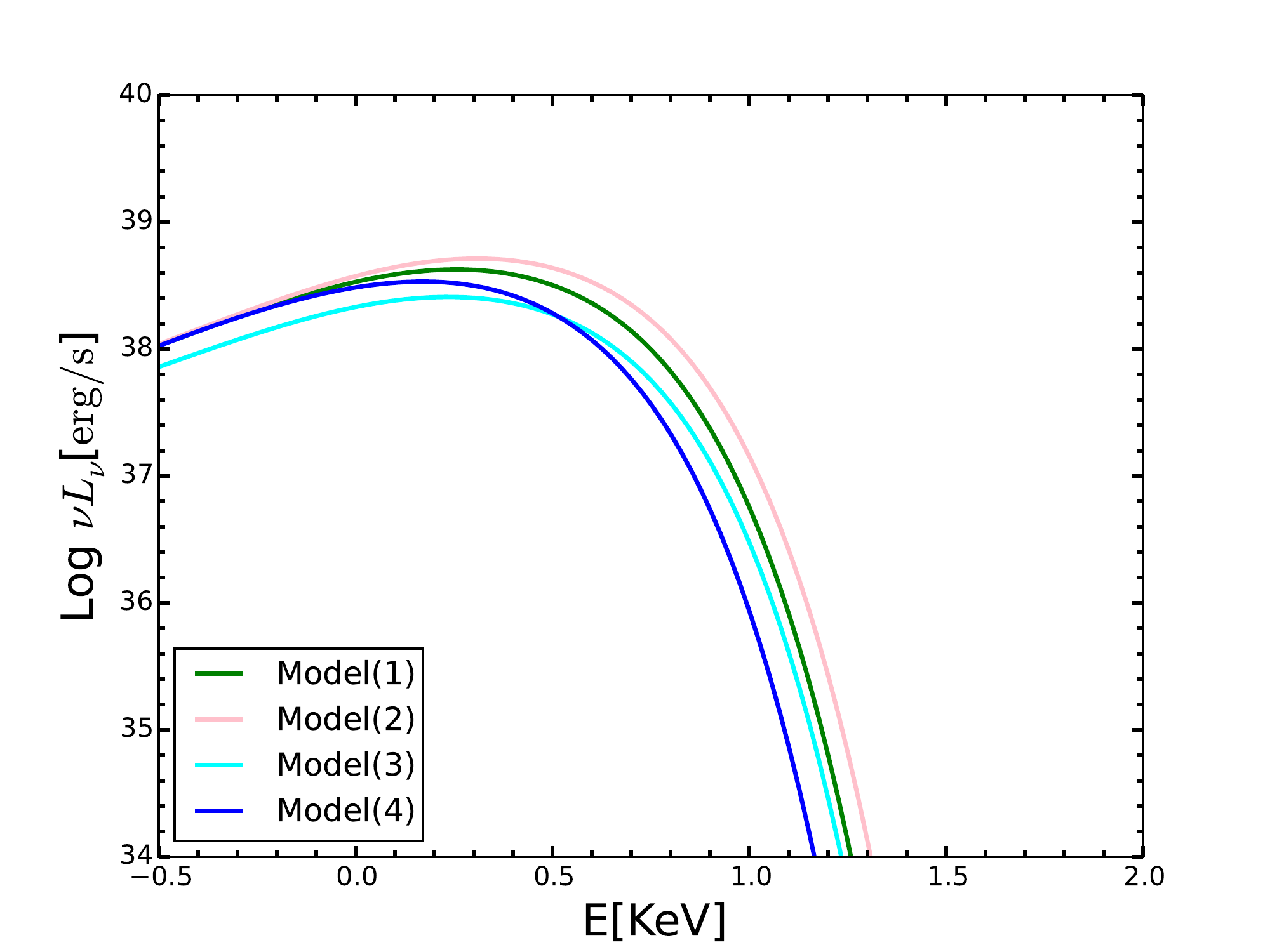}
\caption{
The emergent spectra corresponding to the radial profiles of the accretion rate for the four postulated outflow models in the left panel of Fig. \ref{fig:outflow}.
}
\label{fig:outflow_spectra}
\end{figure}

We define the Eddington limit on the mass accretion rate as $\MdotEdd =  \LEdd/\eta c^2  = 7.57 \times 10^{17} (M / \Msun)$ g/s, where $\eta = 19.01\%$ ($a=0.95$) for GRS~1915+105 and $\MdotEdd =  \LEdd/\eta c^2  = 2.17 \times 10^{18} (M / \Msun)$ g/s, where $\eta = 6.46\%$ ($a=0.2$) for LMC~X-3. With the above definitions of $\MdotEdd$, we have $L/\LEdd = \Mdot/\MdotEdd$ assuming the efficiency is constant. In the following four wind basic models we assume $\Mdot_{\rm out} = 0.8 \ \MdotEdd$ and $r_{\rm out} = 10^6 \ r_{\rm g}$, where $r_{\rm g} = GM/c^2$. The corresponding inflow and mass loss profiles as a function of $r$ are shown in Fig.~\ref{fig:outflow} for $a = 0.95$, while Fig.~\ref{fig:outflow_spectra} illustrates the respective spectra.

\begin{enumerate}
%
\item[(1)]
In the first wind model, we simply assume that the radial distribution of the accretion rate has the form of an exponential function,
\begin{equation}\label{outflow1}
	\Mdot(r) = \Mdot_{\rm out} \exp{ [- \frac{A}{(r/r_{\rm in} - B)}] },
\end{equation}
where $A$ and $B$ are two free variables satisfying $A \geq 0$ and $0 < B < 1$. If there is no wind, $\Mdot(r) = \Mdot_{\rm out}$ which implies $A = 0$. Figure~\ref{fig:outflow} shows the profile for $A = 1.24$, $B = 0.15$, $r_{\rm in} = r_{\rm ISCO}$, and total mass loss rate $\Mdot_{\rm tot} = \Mdot(r_{\rm out}) - \Mdot(r_{\rm in}) \simeq 0.76 \MdotEdd$. 
%
%
\item[(2)] 
In the second model we assume another simple power-law expression for the mass loss rate \citep{2014MNRAS.438.3024L}
\begin{equation}\label{outflow2a}
	\Mdot_{\rm wind}(r) = \Mdot_{\rm out} (r/r_{\rm in})^{\beta}
\end{equation}
which gives the inflow rate 
\begin{equation}\label{outflow2b}
	\Mdot(r) = \Mdot_{\rm out} [ 1 - (r/r_{\rm in})^{\beta}],
\end{equation}
where $\beta < 0$. Figure~\ref{fig:outflow} shows this profile for $\beta = -2.0$, $r_{\rm in} = r_{\rm ISCO}$, and the total mass loss rate $\Mdot_{\rm tot} = 0.8 \MdotEdd$.
%
%
\item[(3)] 
In the third model we adopt the description of the wind suggested by \citet{1982MNRAS.199..883B},
\begin{equation}\label{outflow3a}
	\frac{{\rm d}\Mdot_{\rm wind}(r)}{{\rm d}r} = \frac{C}{r}.
\end{equation}
The continuity equation reads 
\begin{equation}\label{continuity}
	\Mdot(r) = \Mdot_{\rm in} + \int^r_{r_{\rm in}} {\rm d}\Mdot_{\rm wind}(r).
\end{equation}
Integrating Eq.~(\ref{outflow3a}) from $r_{\rm out}$ to $r_{\rm in}$ and comparing it to Eq.~(\ref{continuity}), we derive the inflow rate at a given radius
\begin{equation}\label{outflow3b}
	\Mdot(r) = \Mdot_{\rm in} + \frac{\Mdot_{\rm out} - \Mdot_{\rm in}}{\ln( r_{\rm in}/r_{\rm out} )} \ \ln( r_{\rm in}/r )
\end{equation}
where $\Mdot_{\rm in}$ is a free parameter (see Fig.~\ref{fig:outflow} for $\Mdot_{\rm in} = 0.3 \ \MdotEdd$, $r_{\rm in} = r_{\rm ISCO}$, and total mass loss rate $\Mdot_{\rm tot} = 0.5 \ \MdotEdd$).
%
%
\item[(4)]
Finally, in the fourth model we release the assumption that the disk extends to the ISCO. Instead, we assume that the disk is truncated at $r_{\rm tr}$ due to the wind and that the accretion rate is constant at all disk radii. For simplicity we assume that $r_{\rm tr}$ depends linearly on accretion rate
\begin{equation}\label{outflow4}
	r_{\rm tr} = r_{\rm ISCO} + \Delta r \equiv r_{\rm ISCO} +  D \ \cdot \ \frac{\Mdot_{\rm out}}{\MdotEdd} + E,
\end{equation}
where $r_{\rm tr}$ is in units of $r_{\rm g}$. $D$ and $E$ are two arbitrary free variables (see Fig.~\ref{fig:outflow} for inner truncation at $r_{\rm tr} = 5 \ r_{\rm g}$, and the total mass loss rate $\Mdot_{\rm tot} = 0.8 \ \MdotEdd$). This model may seem particularly unphysical, but it is useful as a parametrization. Its plausible physical interpretation is later discussed in Sec.~\ref{sec:discussion}. 
\end{enumerate}

\begin{figure}[ht!]
\centering
\includegraphics[width=0.45\textwidth]{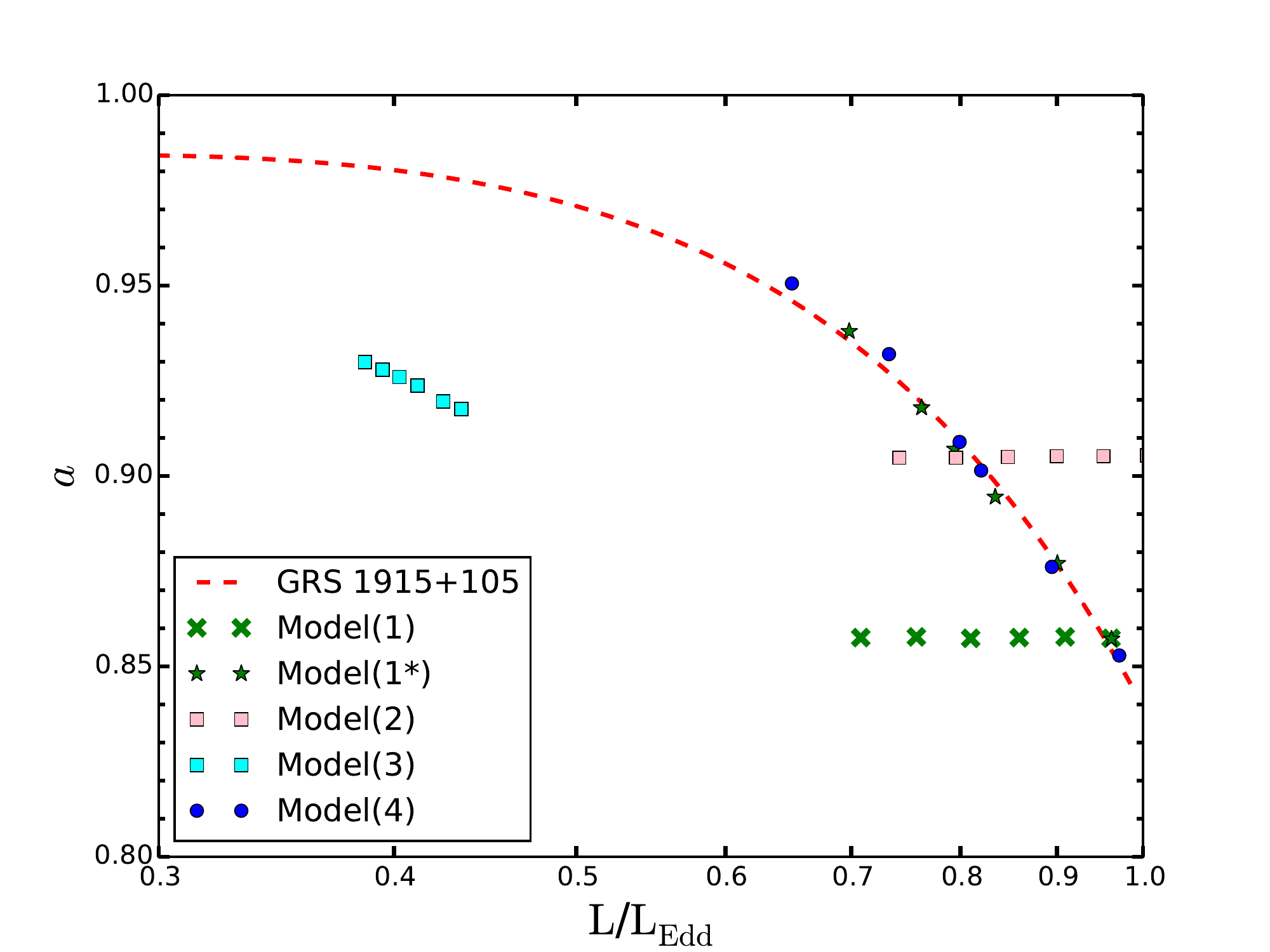}
\caption{
Black hole spin as a function of mass accretion rate in GRS~1915+105. The red dashed line represents a polynomial fit to the spin estimates in our companion paper (Straub et al., in preparation) derived from fitting the slim disk model against the data set of \cite{2006ApJ...652..518M} with the dynamical parameters \citep{2014ApJ...796....2R} to date. The spectra of our four outflow models are calculated for a range of $\Mdot_{\rm out}$ and a constant black hole spin $a = 0.95$. The green crosses, pink squares, cyan squares, and blue circles show the points ($\Mdot_{\rm out}^{\prime}$, $a^{\prime}$) obtained from fitting our outflow model spectra with relativistic standard model \citep{1973blho.conf..343N}. The green stars are the estimates based on Model~(1) assuming parameter $A$ depends on mass accretion rate.}
\label{fig:1915}
\end{figure}

\begin{figure}[ht!]
\centering
\includegraphics[width=0.45\textwidth]{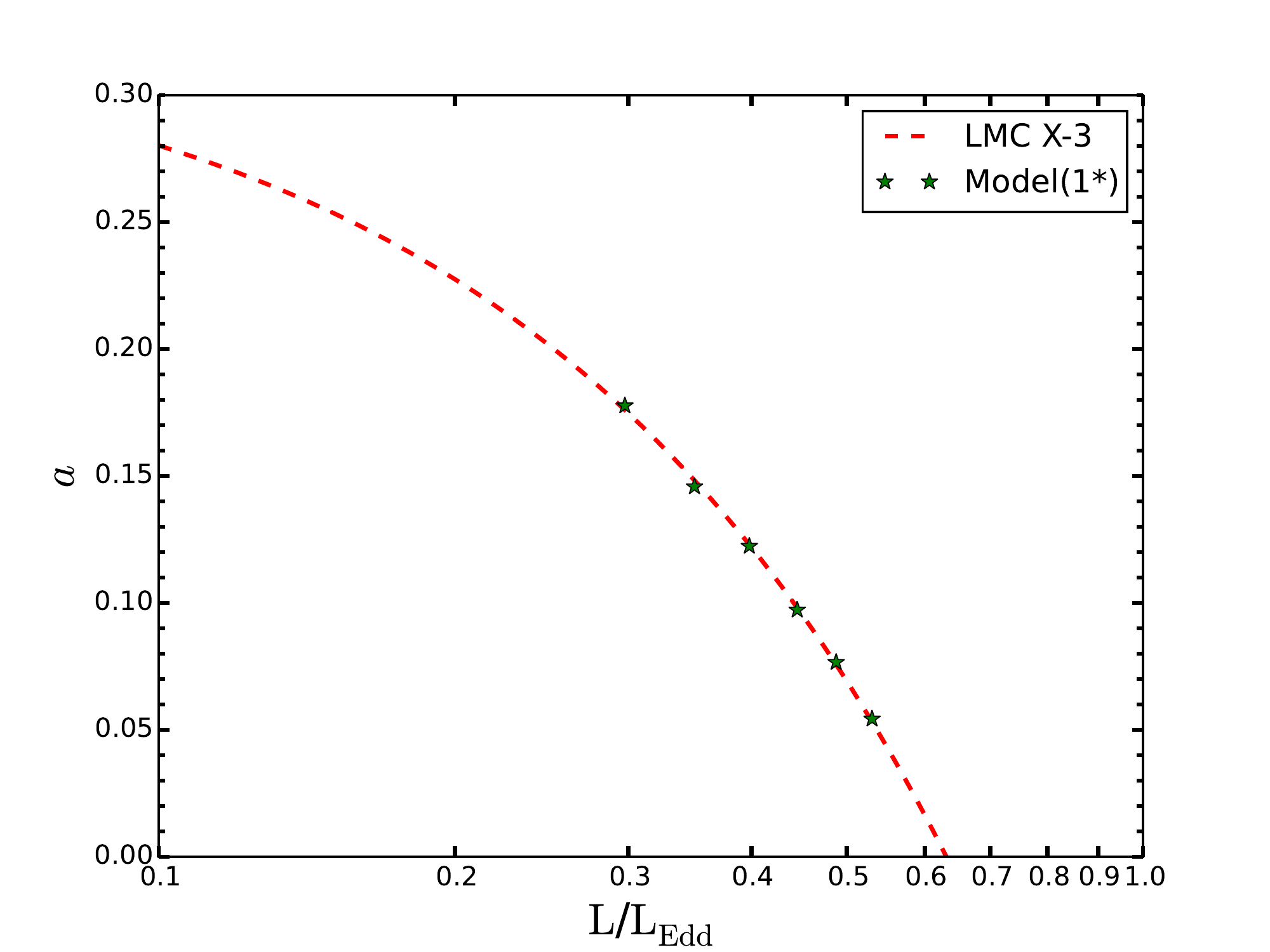}
\caption{
Black hole spin as a function of mass accretion rate in LMC~X-3. The red dashed line represents a linear fit to the spin estimates in our companion paper (Straub et al., in preparation) derived from fitting the slim disk model against the data set of \cite{2010ApJ...718L.117S} with the dynamical parameters \citep{2014ApJ...794..154O} to date. The green stars are obtained by fitting our outflow Model~($1^{*}$) with the Novikov-Thorne disk model.
}
\label{fig:lmcx3}
\end{figure}

We now try to recreate the measured accretion rate dependency of the black hole spin by fitting disk spectra based on the four wind models above with a standard wind-less thin disk model \citep{1973blho.conf..343N}. 
This part of analysis does not involve any real spectral data but is based on comparison of models with and without wind. We use the Fortran ray-tracing code described in \citet{2011MNRAS.415.2942C} which includes all general relativistic effects. The local emissivity follows the Novikov-Thorne prescription but the local accretion rate decreases inwards with radius in the presence of winds. The disk spectra are parametrized by input mass accretion rate and spin ($\Mdot_{\rm out}$, $a$) for an assumed viewing angle, $i$, plus the parameters of the individual wind model. We calculate the resulting spectra including light bending effects. In the next step we treat these synthetic spectra as ``observed spectra'' and apply a maximum likelihood method to fit them (still using Fortran) with the original wind-less Novikov-Thorne prescription for the local flux. We compare the two models in the range $\sim$ 0.1-100.0 keV. The accretion rate $\Mdot_{\rm out}^{\prime}$ and the spin $a^{\prime}$ are now the fit parameters. These are plotted in Fig.~\ref{fig:1915} and Fig.~\ref{fig:lmcx3} for GRS~1915+105 and LMC~X-3, respectively. In both figures the red dashed line indicates the best linear fit to the bulk of spin measurements obtained from fitting the slim disk model to the actual source spectra. The question is whether the derived sets of ($\Mdot_{\rm out}^{\prime}$, $a^{\prime}$) from fitting the synthetic spectra follow these reference lines from fits to real spectra. There are no observational errors or instrument sensitivity effects, so the absolute values of $\chi^2$ are meaningless but the procedure allows to find the best match between the models with and without wind.

The radial mass loss profiles presented in Fig.~\ref{fig:outflow} imply that Model~(2) shows the steepest dependence on the radius. However, the inflow rate only significantly decreases in a small portion of the disk close to ISCO and has little influence on the shape of the model spectrum compared to other models (see Fig.~\ref{fig:outflow_spectra}). The radial change of the accretion rate in Model~(1) is in general steeper than that in Model~(3) at the inner radius. This means that Model~(1) will affect the disk spectrum, especially the contribution from the inner disk radius, more significantly which makes it more interesting for our case. However, extending the analysis to spectra from $L/L_{\rm Edd} \simeq 1.0$ to $L/L_{\rm Edd} \simeq 0.7$ we notice that Model~(1) overproduces the mass loss rate which results in much lower black hole spins (green crosses in Fig.~\ref{fig:1915}) compared to the measured ones, since the character of the current radial change of the accretion rate is not coupled strongly enough with the accretion rate. The $\Mdot_{\rm out}$ - dependent Model~(4) is the simplest possible model that is able to reproduce the observed apparent spin decrease. In the parameter space (D, E) when simulating the decline of spin with mass accretion rate, $D = 8.55^{+1.71}_{-3.35}$ and $E = 5.00^{+1.53}_{-2.80}$ give the best fits (the blue circles in Fig.~\ref{fig:1915}) to the red dashed line. The uncertainties are quoted at the 90\% confidence level by assuming an error of $\Delta a = 0.01$ in the spin of the red dashed line. Interestingly, we can achieve an equally good match with Model~(1), by assuming that parameter $A$ depends linearly on the mass accretion rate, i.e. Model~(1*), $A(\Mdot_{\rm out}) = A_{1} \cdot \Mdot_{\rm out}/\MdotEdd + A_{2}$. On the one hand, this means that when no winds are present, the exponential function in Eq.~\ref{outflow1} must equal 1 and thus $A(\Mdot_{\rm out}) = 0$. On the other hand, an increasing $A(\Mdot_{\rm out})$ reflects an increasing contribution of winds that reduces the local mass accretion rate and modifies the spectrum. This works because a smaller $A$ at low mass accretion rates gives a shallower shape to the radial distribution of the accretion rate. In this manner the $\Mdot_{\rm out}$ - dependent Model~($1^{*}$) reduces the influence of the wind on the Novikov-Thorne spectrum at low accretion rates. 

When simulating the behavior of $a$ versus $\Mdot_{\rm out}$ for GRS~1915+105 in the range $L > 0.6 L_{\rm Edd}$ with Model~($1^{*}$) we explore the parameter space ($A_{1}, A_{2}$). Parameter B is arbitrarily fixed at B = 0.15 since the radial distribution of the accretion rate is insensitive to it. We find that $A_{1} = 3.98^{+1.42}_{-1.25}$ and $A_{2} = -2.54^{+1.02}_{-1.03}$ give the best fits to the observed spin trend (see the green stars on the red dashed line in Fig.~\ref{fig:1915}). 
We repeat the same analysis for LMC~X-3 in the range of $0.3 L_{\rm Edd} < L < 0.6 L_{\rm Edd}$. We find again that Model~($1^{*}$) can describe very well the observed spin trend with $A = 1.57^{+0.40}_{-0.25}  \cdot  \Mdot_{\rm out}/\MdotEdd - 0.39^{+0.17}_{-0.10}$. The reconstructed values of ($\Mdot_{\rm out}^{\prime}$, $a^{\prime}$) are plotted in Fig.~\ref{fig:lmcx3} as green stars. 
The use of fixed values A1 and A2 works only in a limited range of accretion rates but this simple exercise shows that the requested wind is highly concentrated towards the ISCO: in GRS 1915+105 more than 50 per cent of wind mass loss takes place below 5 $r_{\rm g}$ and in LMC X-3 most of the wind occurs inside 14 $r_{\rm g}$. Thus, in both cases the radial range for the apparent wind is only about three times the size of the ISCO. 

From a cross-correlation analysis of optical/infrared (OIR) data and contemporaneous X-ray data, \citet{2014ApJ...783..101S} found that there is a non-linear relation between the OIR flux and the time-lagged X-ray emission: $F_{\rm OIR} \propto F_{X}^{\beta}$ with $\beta \simeq 1.3$. This may indicate that the rates of inflow at the inner and outer radii are not matched and that the ratio of the inflow rate scales inversely with the OIR flux, i.e. $\Mdot_{\rm in}/\Mdot_{\rm out} \propto \Mdot_{\rm out}^{-3/13}$. Disk winds are suggested to carry the accreted gas away from the disk. Given the above $\Mdot_{\rm out}$ - dependence on the parameter $A$ and Eq.~\ref{outflow1}, our Model~($1^{*}$) predicts that the ratio of $\Mdot_{\rm in}/\Mdot_{\rm out} \propto exp(-\Mdot_{\rm out})$, which is fully consistent with the anti-correlation between the inflow rate and $\Mdot_{\rm out}$ found by \citet{2014ApJ...783..101S}.

Our simple parametric discussion above shows that Model~($1^{*}$), Model~(4), or possibly more sophisticated models not considered here can describe such a mass loss and successfully reproduce the observed decrease in spin over a range of mass accretion rates. The removed material will be present between the disk and the observer and may contribute to additional modifications to the observed spectrum.

\section{Physics of wind and Comptonization of the disk spectrum by the surrounding material}
\label{sec:comp}
Our simple parametric description of the thermal emission from an accretion disk with mass losses used in Sec.~\ref{sec:flow} is too simplistic to predict the resulting effects of the removed material on the observed spectra. 

The most general description of a 1-D disk structure with the wind should include both the mass loss from the disk, implying the dependence of the accretion rate on the radius, $\Mdot(r)$, as well as the energy extraction by the wind. Thus, in general, the local radiation flux, $F(r)$, in the disk is given by
\begin{equation}
	F(r) = {3 G M \dot M \over 8\pi r^3} {\cal L}(r)(1 - f_{\rm wind})
\end{equation}
if the particles in the wind carry more energy than the remaining particles in the disk. Here ${\cal L}$(r) describes the standard term for relativistic effects in the Novikov-Thorne disk, and the energy fraction $f_{\rm wind}$ accounts for additional removal of the energy by the wind.  We need to specify two arbitrary radial functions, $\dot M(r)$ and $f_{\rm wind}$, to fully account for the impact of the wind on the disk structure. The relative importance of the two effects likely depends on the wind driving mechanism. 

There are a number of wind mechanisms proposed which may act on a cold accretion disk surface: (i) thermally driven winds \citep{1958ApJ...128..664P,1993ApJ...402..109B,1996ApJ...461..767W}; (ii) magnetically driven winds \citep{1982MNRAS.199..883B,1994ApJ...434..446K,1994ApJ...429..139C}; (iii) radiation pressure driven winds, which in turn include Compton-driven and line-driven winds \citep{1985ApJ...294...96S,2000ApJ...543..686P,2014ApJ...789...19H}. Thermally driven winds and radiation pressure driven winds do not require considerable energy segregation at the launch radius and in this case the role of the $f_{\rm wind}$ factor can be possibly neglected although outflowing plasma needs extra energy to achieve escape velocity. In a Newtonian Keplerian motion, it requires doubling the test particle energy to unbind it. Thus the effect is small as long as the ratio of the total wind mass flux and the accreted mass flux is not too high. On the other hand, most energy and angular momentum can be efficiently extracted from the accretion flow by a small fraction of the magnetically driven plasma (winds or jets).

In our description of the disk in Sec.~\ref{sec:flow} we assumed $f_{\rm wind} = 0$ for simplicity. If the wind is actually a thermally driven wind, our expression for the mass loss indeed implies the true mass loss from the system. If the magnetic mechanism dominates, our simplistic method gives an upper limit on the true mass loss from observational constraints. 

In any of these cases the material may, or may not, be present along the line of sight since this will depend of the wind geometry for a given inclination of an observer. If the wind is mostly collimated, the line of sight may miss most of the removed material, but if the wind is roughly spherical (or, accidentally, mostly towards an observer) we should see signatures of this wind in the form of Comptonization effect and/or absorption lines.

An example of a continuum radiation driven wind has been discussed in detail by \citet{2003MNRAS.345..657K} \cite*[also see][]{2016MNRAS.455.1211K}. Such an outflow is expected in sources radiating close or above the Eddington limit. If most of the material does not reach the ISCO but flows out, it may form a roughly spherical outflow or a collimated wind in a double cone. The analytic solutions for both geometries showed that the outflow may be Compton-thick and is a promising candidate for the soft excess observed in many AGNs and ultraluminous X-ray sources, given that much of the accretion energy must emerge as blackbody emission due to its large optical depth. The parameters of such an outflow were discussed by \citet{2003MNRAS.345..657K}. If the wind is spherical, its optical depth can be described as
\begin{equation}\label{eq:wind}
	\tau_{\rm wind} \propto \frac{1}{\eta} \frac{r_{\rm g}}{r_{\rm in}} \frac{\Mdot_{\rm tot}}{\MdotEdd} \frac{c}{v},
\end{equation}
where $\Mdot_{\rm tot}$ is the total mass outflow rate, and $v$ is the wind velocity which is higher than the escape velocity from a given radius. For sources that accrete close to the Eddington limit this optical depth is relatively large.  However, the temperature, $T_{\rm wind}$, of such an outflow is not very high since there is not enough additional energy to heat the plasma. So, the temperature of the outflow stays close to the disk temperature. Another possibility which is also discussed in \citet{2003MNRAS.345..657K} is that the outflow is highly collimated and of a jet-like form. 

\begin{figure}[ht!]
\centering
\includegraphics[width=0.5\textwidth]{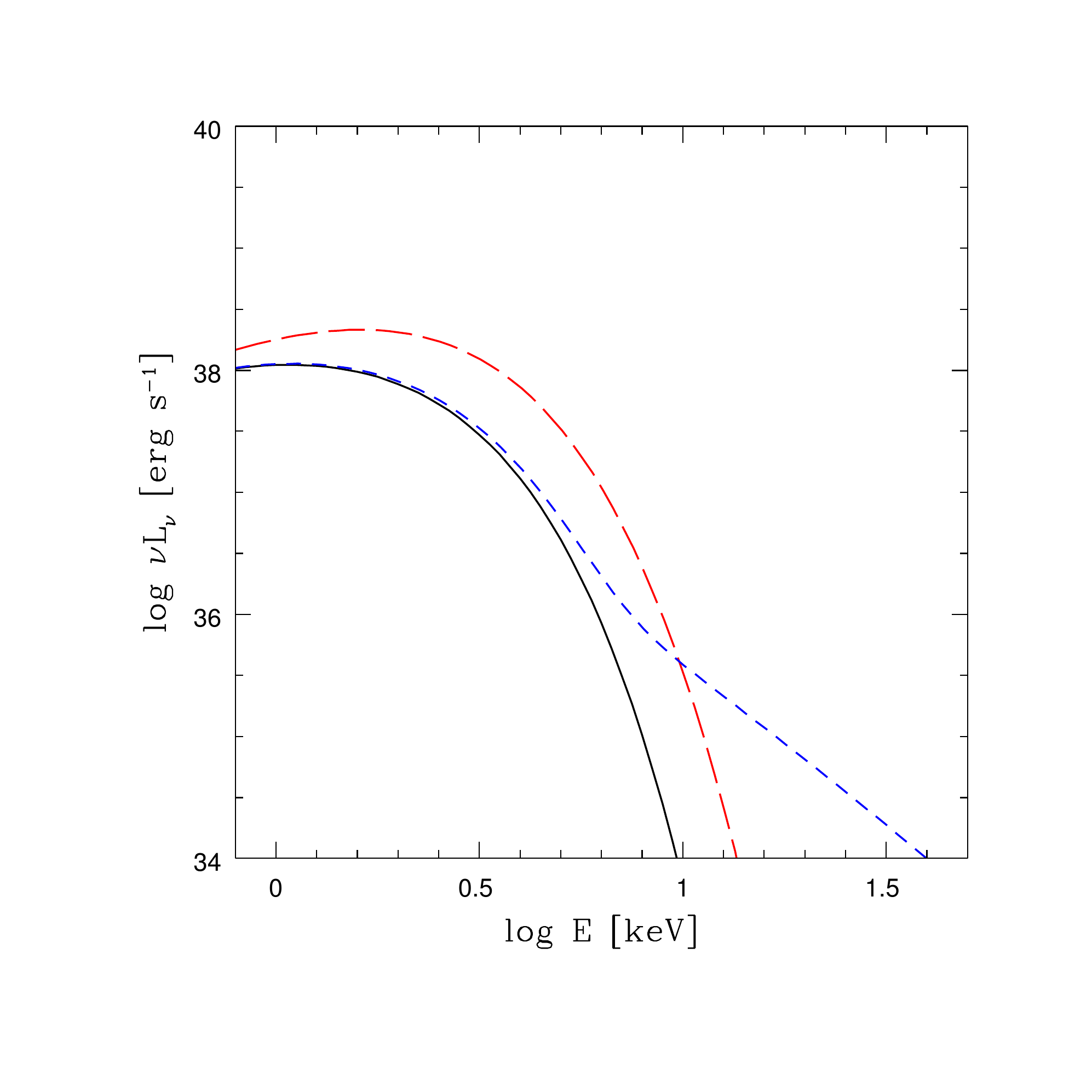}
\caption{
Example of the disk radiation spectrum with outflow as in Model~($1^{*}$), {\it before} passing through a spherical wind layer (black solid line) and {\it after} Comptonization by a spherical wind with (i) $\tau = 5$ and $T = 1 $keV (red long-dashed line) or (ii) $\tau = 0.05$ and $T = 100 $keV (blue short-dashed line).
}
\label{fig:spectra1}
\end{figure}

In Fig.~\ref{fig:spectra1} we show the predicted disk spectrum seen through such a cold optically thick emerging wind for GRS~1915+105 predicted by Model~($1^{*}$, red long-dashed line), and  compare it to the initial Model~($1^{*}$) spectrum (black solid line). A cold wind softens the low-energy part of the spectrum and hardens the spectral peak and the high-energy part, and does not produce a hard X-ray power law tail. Since the wind density is relatively high, we have to estimate the possible role of the bremsstrahlung radiation. If the outflow consists of fully ionized pure hydrogen gas, the luminosity of the bremsstrahlung radiation is 
\begin{equation}
\begin{aligned}
	L_{\rm brem} ={} & 1.68 \times 10^{-27} \ T^{1/2} \  \int^{\infty}_{r_{\rm in}}n^2(r)4\pi r^2dr \\
	& = 1.68 \times 10^{-27} \ T^{1/2} \ \frac{4 \pi r_{\rm in} \ \tau^2 (1-\beta)^2}{\sigma_{T}^2 \ (2\beta -3)} \quad  \rm erg/s
\end{aligned}	
\end{equation} 
where the density of the wind is assumed to be $n(r) = n_{0} (r/r_{\rm in})^{-\beta}$, and then the optical depth $\tau = \int^{\infty}_{r_{\rm in}} n(r) \sigma_{T}dr = n_{0}\sigma_{T}r_{\rm in}/(\beta-1)$. For $T = 1$keV, $\tau = 5$, $\beta = 2$, and $r_{\rm in} = 1.94 \ r_{\rm g}$ (at $a = 0.95$) one gets $L_{\rm brem} = 1.44 \times 10^{34}$ erg/s, while the luminosity of the outflow spectra (black solid line) is $L_{\rm disk} = 2.29 \times 10^{38}$ erg/s. Therefore, the emergent spectra from the outflow are dominated by the disk emission and the Comptonization, and the model neglecting the bremsstrahlung is self-consistent. Thus the bremsstrahlung radiation will not be taken into account in further considerations. 

On the other hand, the emerging wind may be magnetic in nature, or a weak static magnetic corona can be present above the accretion disk, as frequently postulated for soft states \citep{2007A&ARv..15....1D, 2009MNRAS.394..207C, 2012ApJ...761..109Y}. We then have no particular predictions of the properties of the Comptonizing medium from our model. Therefore, we arbitrarily assume the values of $T = 100$ keV, $\tau = 0.05$,
which implies a small fraction of energy in this component (Comptonization parameter $y = 0.07$). We show this example for the same disk model in Fig.~\ref{fig:spectra1} (blue short-dashed line). A hot Comptonizing medium marginally hardens the spectral peak and produces a hard high-energy tail to the thermal component which is in general rather steep due to the efficient supply of the soft photon from the disk and the negligence of the outflow velocity.  

\begin{figure}[ht!]
\centering
\includegraphics[width=0.5\textwidth]{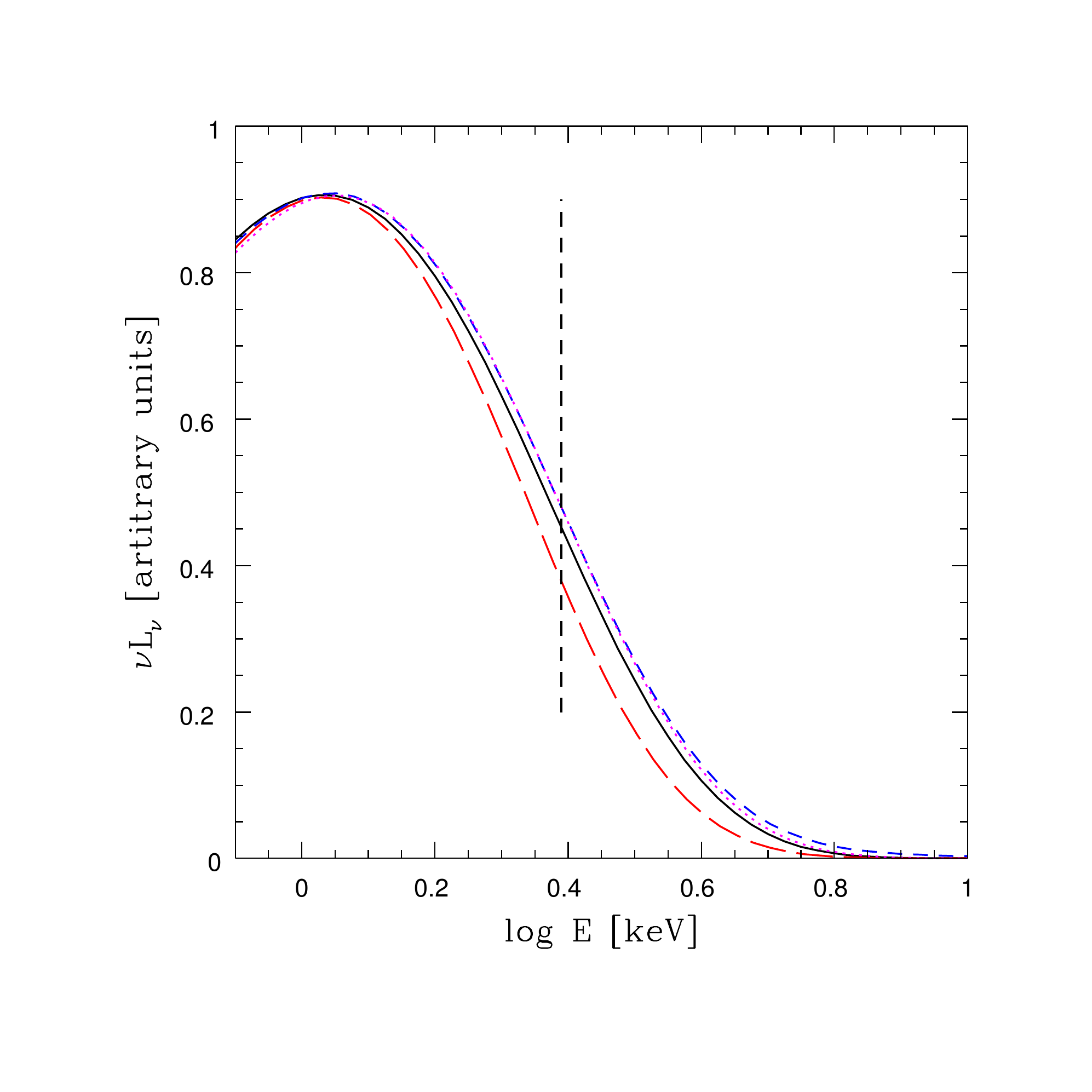}
\caption{
Shape of the thermal component normalized to the peak position shown in a linear scale for a relativistic disk without outflow (magenta dotted line), a disk with outflow as in Model~($1^{*}$) but before passing through a spherical wind layer(black solid line), and after Comptonization by a spherical wind with (i) $\tau = 5$ and $T = 1 $ keV (red long-dashed line) or (ii) $\tau = 0.05$ and $T = 100 $keV (blue short-dashed line). The vertical dashed line indicates the energy at which the flux of the outflow disk spectrum before Comptonization (black solid line) is factor 2 below the peak.
}
\label{fig:spectra2}
\end{figure}

In order to show the magnitude of the effect of Comptonization inside the wind more clearly, we plot the renormalized and shifted spectra in a linear scale, with peaks at the same energy, in Fig.~\ref{fig:spectra2}. We can now compare them at the energy where the initial Model~($1^{*}$) disk flux is a factor 2 below its peak flux. At this location the differences between the models should show more clearly in the data, since at higher energies the data usually requires additional hard X-ray component. We see that  the spectrum where photons go through a cold optically thick wind is visibly steeper with fewer photons at the given energy, while the spectrum where photons go through a hot optically thin wind is less steep with slightly elevated flux. For comparison, we also added the Novikov-Thorne model without any form of outflow (magenta dotted line). This spectrum is also slightly harder than Model~($1^{*}$) and the cold wind case. The difference between the two disk models with/without outflows is at the level of $\sim4\%$, again at the energy where the flux of the disk before Comptonization is a factor 2 below the peak (see Fig.~\ref{fig:spectra2}). However, the difference between the disk without wind and the disk with wind and optically thin Comptonization (i.e. between magenta dotted line and the blue short-dashed line) is very small, only $\sim0.2\%$, almost invisible in the figure scale. Observational data of good quality could thus allow us to discriminate (within the observational error) between the different spectral shapes shown in Fig.~\ref{fig:spectra2} and enable us to assess not only whether wind resolves the spin problem in accretion disks, but also which of the outflow/wind scenarios presented above is the preferred one.

\begin{figure}[ht!]
\centering
\includegraphics[width=0.45\textwidth]{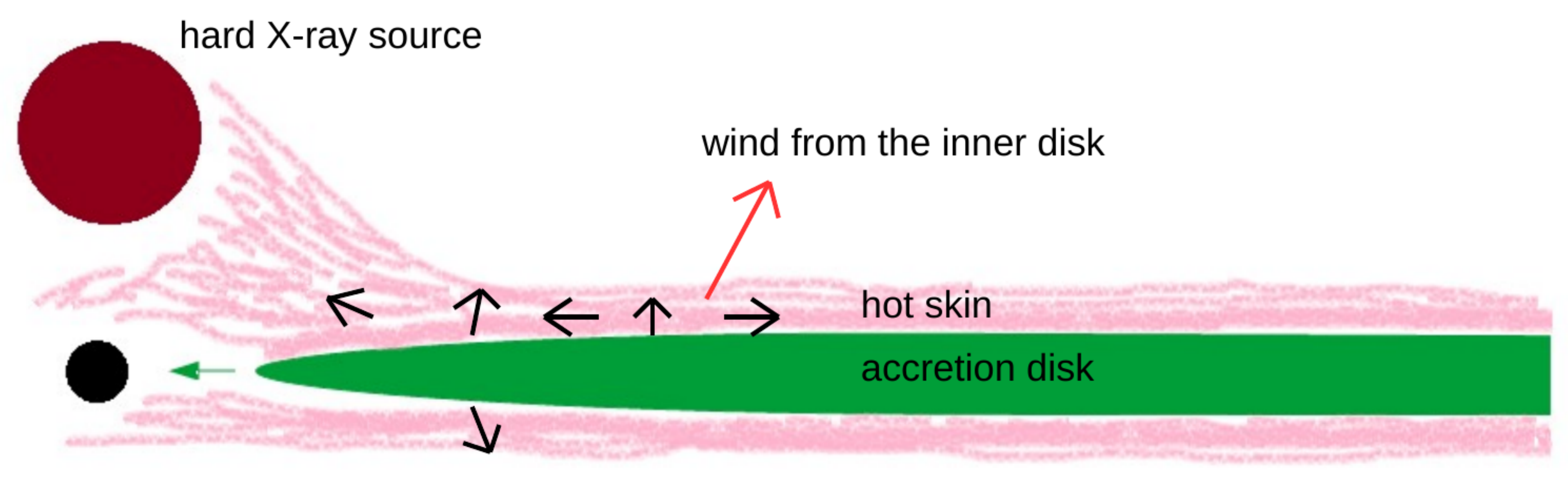}
\caption{
The sketch of the innermost part of the accretion flow for moderate/high Eddington ratio: the disk extends to ISCO, and it is surrounded by the hot skin which likely forms a sub-Keplerian uncollimated wind, and the hard X-ray source which is most likely the base of the jet. 
}
\label{fig:sketch}
\end{figure}

Simple models discussed above are order of magnitude estimates and do not rely on detailed physical studies; they consider only the simplest geometry which almost certainly differs from real accreting systems. Recent modeling of hard X-ray emission in soft state XRBs and AGN with Nuclear Spectroscopic Telescope Array ({\it NuSTAR}) indicates that the emission comes from a very compact region, 3 to 10 $r_{\rm g}$ in size, that is well approximated by a lamp-post located on the symmetry axis \citep[e.g.][]{2015MNRAS.451.4375F}. This region can be identified as the jet base. Although the jet seems to be suppressed in the high/soft state in XRBs \citep{1999ApJ...519L.165F, 2004MNRAS.355.1105F}, the jet base may remain present and be related to uncollimated winds and significant energy dissipation near the black hole. The material presumably comes from an accreted hot corona as it has been discussed in a number of papers over the past decade \citep[e.g.][]{1995ApJ...455..623C, 1995MNRAS.277...70Z, 2015ApJ...806..223L, 2015MNRAS.449..129W}. Since winds due to radiative or magnetic acceleration from the inner disk region must be Compton-thin as demonstrated by \citet{2012ApJ...759L..15R}, the general scenario of a cool, optically thick disk wrapped in a hot, optically thin corona, where a fraction of the coronal material is removed in form of winds while the rest is accreted, applies. We sketch this in Fig.~\ref{fig:sketch} (see the discussion in the following section).
A detailed computation of the emission from such a complex model is beyond the scope of the present paper as this would require knowledge of the dissipation in the coronal hot flow. However, we can verify if the simple wind model discussed in Sec.~\ref{sec:flow} is self-consistent and determine which part of the flow is actually seen. 
We present such observational tests in the following Sec.~\ref{sec:tests} for LMC~X-3 and GRS~1915+105 using Model ($1^*$).

\section{Testing the wind scenario in LMC X-3 and GRS 1915+105}
\label{sec:tests}
In the previous section it was shown that in principle one can test the wind scenario if the observational data determine not just the position of the spectral peak and the normalization but if they are also sensitive to the precise shape of the thermal component. In this section we fit Model~($1^{*}$) against observational data of LMC~X-3 and GRS~1915+105. The spectral analysis on all data sets is performed with the X-ray spectral fitting software package, XSPEC v.12.8.2. In order to implement our disk-wind model into the framework of XSPEC we construct a table model {\sc diskw} for which each individual spectrum is calculated with a modified version of the code by \citet{2011MNRAS.415.2942C} based on \citet{1973blho.conf..343N}, which accounts for the modified accretion rate given in Model~($1^{*}$). The description of the hardening factor adopted in the model spectrum is done through the parametrization of {\sc bhspec} results for the viscosity parameter $\alpha = 0.1$
\[
f_h =
  \begin{cases}
   1.6 \ [(\dot m + 0.1)/0.2]^{0.24}                & \text{for} ~~~ T_4 > 10 \\
   (T_4/3.0)^{0.3904} \ [(\dot m + 0.1)/0.2]^{0.24} & \text{for} ~~~ 1< T_4 < 10 \\
   [(\dot m + 0.1)/0.2]^{0.24}                      & \text{for} ~~~ T_4 < 1,
  \end{cases}
\]
where $\dot m = \Mdot/\MdotEdd$ and $T_4 = T/10^4$ K. It partially comes from \citet{2012MNRAS.420.1848D} and Eq.~(A13) in \citet{2006ApJ...647..525D}, but includes the explicit term in accretion rate which is absent in those papers. The hardening factor increases monotonically with temperature. Therefore, in our table model, there are only two free parameters, parameter $A$ (see Eq.~\ref{outflow1}) and the initial mass accretion rate $\Mdot_{\rm out}$ at the outer disk edge, i.e. before the disk is affected by mass loss due to the wind. The mass accretion rate, $0.15 \leq \dot m \leq 1.5$, is scaled by the Eddington limit of a given source, $\dot m = \Mdot_{\rm out}/\MdotEdd$ (see Sec.~\ref{sec:flow} for the definition of $\MdotEdd$). 
Unlike in the initial Model ($1^*$) in Sec.~\ref{sec:flow}, the free input parameters $A$ and $\dot m$ are not coupled in the fitting here. 
Parameter $B$ is fixed at $B = 0.15$, since the radial distribution of the accretion rate is insensitive to it. For each source we use the respective up-to-date binary parameters given in Sec.~\ref{sec:estim} and the resulting mean spin values derived from the spectra obtained in low disk luminosities, $L \leq 0.3 \LEdd$, in fits with the slim disk model (Straub et al., in preparation). Our base model to fit the soft state spectra of both sources reads 
\begin{center}
	{\sc tbabs * (diskw + nthcomp)}.
\end{center} 
The component {\sc tbabs} \citep{2000ApJ...542..914W} accounts for photon absorption by neutral hydrogen in the direction of the source. The column density towards LMC~X-3 is $N_H = 4 \times 10^{20}\,$ cm$^{-2}$ \citep{2003MNRAS.345..639P} and towards GRS~1915+105 it is $N_H = 5 \times 10^{22}\,$ cm$^{-2}$ \citep{2002ApJ...567.1102L}. The disk component is normalized by a constant factor 0.0575 which converts the units between the table model and the energy bins. Both sources require an additional component to account for the high energy photons. We use the thermal Comptonization model {\sc nthcomp} \citep{1996MNRAS.283..193Z, 1999MNRAS.309..561Z}. The photon counts at high spectral energies have fairly large uncertainties, so in order to avoid degeneracies with other parameters the photon index which defines the powerlaw slope is allowed to vary only in the range $1.5 < \Gamma < 3.5$. The electron temperature, $kT_{\rm e}$ and the normalization are left free, whereas the seed photon temperature, $kT_{\rm bb}$, is fixed. We get lower limit on $kT_{\rm bb}$ by first running fits with the non-relativistic disk model {\sc diskbb} \citep{1984PASJ...36..741M} which are found to be reasonable values as for the temperatures of the disk with wind, and those inner disk temperatures are used as fixed seed photon temperatures in {\sc nthcomp}.  

In the spectral fits with {\sc diskw} the vertical structure of the disk is not taken into account. In particular the disc thickness is assumed to be small so that the photons are emitted from the equatorial plane, which is most likely inappropriate for high mass accretion rates.

\begin{deluxetable}{lcccc}
\tablecaption{Observation dates given in MJD.\label{tab:obsid}} 
  \tablehead{
  \colhead{source} & \colhead{Obs 1} & \colhead{Obs 2} & \colhead{Obs 3} & \colhead{Obs 4}
  }
  \startdata
LMC~X-3 & 53607 & 52138 & 51172 & 52002 \\
GRS~1915+105 & 50763 & 50756 & 50190 & 52996 
\enddata

\end{deluxetable}

\subsection{Observational tests with LMC~X-3}
\begin{figure}[ht!]
\centering
\includegraphics[width=0.45\textwidth]{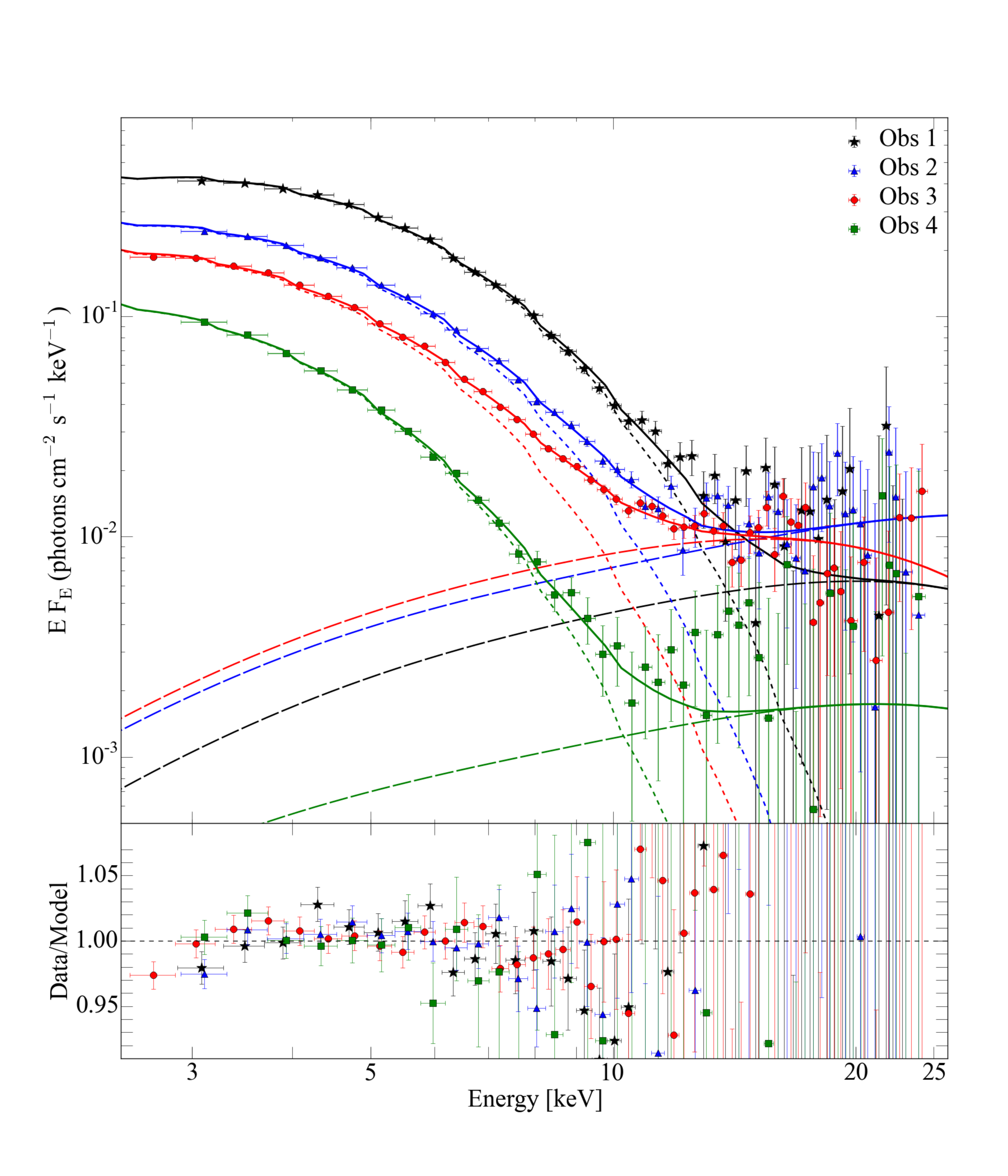}
\caption{
Four RXTE soft state spectra of LMC~X-3 with mass accretion rates between $\approx 0.16 - 1.26 \ \MdotEdd$, fitted individually over the energy range 2.51 - 25.0 keV. The full model (solid lines) is composed of an absorbed outflow-disk component (short dashed lines) and a Compton component (long dashed lines). The fit results are shown in Tab.~\ref{tab:lmcx3_result}.}
\label{fig:lmcx3_fit}
\end{figure}

We look at the sample of 712 X-ray spectra that have been obtained using the large-area Proportional Counter Array (PCA) aboard the {\it Rossi X-ray Timing Explorer} (RXTE). These data were extracted from the Proportional Counter Unit 2 (PCU-2), background subtracted and corrected for a detector dead time of $\sim 1 \%$, and classified as the soft states \citep[a detailed description is given in][]{2010ApJ...718L.117S, 2014ApJ...793L..29S}. We fit these soft state spectra with the above given base model {\sc tbabs * (diskw + nthcomp)} over the energy range 2.51-25.0 keV. A good fit satisfies (i) $\chi^2_{\nu} \sim 1$ and (ii) the unabsorbed disks flux must be at least $75 \%$ of the total unabsorbed flux. Out of the total number of soft state observation 381 spectra have mass accretion rates in our range of interest $(\dot m \geq 0.15)$, where winds are expected to play a role. We obtain 336/381 good fits. Parameter $A$ decreases monotonically with mass accretion rate from about 4.5-0.03 while $\dot m$ decreases from 1.5-0.15 (see Fig.~\ref{fig:amdot}). The behavior of parameter $A$ suggests that the wind in this source decreases markedly, but never entirely dies down. The estimated seed photon temperatures take values $kT_{\rm bb} \approx 1.5 - 0.9$ keV and the electron temperatures lie between $kT_{\rm e} \approx 20 - 3.5$ keV. From the goodness of fit and the residuals we see that a disk-wind model based on a fixed spin, $a = 0.20$, represents the LMC~X-3 data well. We choose four representative spectra in a wide range of mass accretion rates and show their data, the models, and the residuals in Fig.~\ref{fig:lmcx3_fit}. The results are summarized in Table~\ref{tab:lmcx3_result}.

\begin{deluxetable*}{lcccc}[!t]
\tabletypesize{\scriptsize} \tablecaption{{\sc diskw} fits to LMC~X-3 soft state spectra\label{tab:lmcx3_result}} \tablewidth{0pt} 
\tablehead{ \colhead{model} &
\colhead{Obs 1} & \colhead{Obs 2} & \colhead{Obs 3} & \colhead{Obs 4} } \startdata
{\sc diskw} &&&& \\ [2pt]
%
$\Mdot_{\rm out}/\MdotEdd$ & 
$1.26\ ^{+0.08}_{-0.08}$ &
$0.53\ ^{+0.03}_{-0.03}$ &
$0.33\ ^{+0.02}_{-0.02}$ & 
$0.16\ ^{+0.02}_{-0.02}$  \\ [5pt]
$A$ & 
$3.86\ ^{+0.35}_{-0.35}$ &
$1.70\ ^{+0.24}_{-0.24}$ & 
$0.87\ ^{+0.16}_{-0.16}$ & 
$0.36\ ^{+0.35}_{-0.14}$ \\ [5pt] 
\hline
{\sc nthcomp} &&& \\[2pt]
%
$\Gamma$ & 
$1.5\ ^{+0.81}_{*}$ &
$1.5\ ^{+0.2}_{*}$ & 
$1.5\ ^{+0.21}_{*}$ & 
$1.5\ ^{+0.48}_{*}$ \\[5pt]
$kT_{\rm e}$ [keV] & 
$9.57\ ^{*}_{-3.00}$ &
$3.87\ ^{+0.96}_{-1.05}$ & 
$5.03\ ^{+2.31}_{-0.77}$ & 
$5.80\ ^{*}_{-2.95}$  \\[5pt] 
$kT_{\rm bb}$ [keV] & 
1.26 &
1.17 & 
1.13 & 
0.92 \\[5pt] 
N [$10^{-4}$] & 
$1.10\ ^{+2.36}_{-0.31}$ &
$1.81\ ^{+0.10}_{-0.24}$ &
$2.01\ ^{+0.98}_{-0.12}$ &
$0.59\ ^{+2.04}_{-0.17}$ \\[5pt] 
\hline \\ [1pt]
$\chi^2$/dof ($\chi^2_{\nu}$) & 
50.24/45 (1.12) &
45.03/44 (1.02) & 
42.68/49 (0.87) &
21.32/44 (0.49) \\[5pt] 
%
%
\enddata
\tablecomments{The best fitting spectral parameters. All errors are quoted at the $90 \%$ confidence level ($\Delta \chi^2 = 2.706$). The asterisk, $ ^*$, indicates an unconstrained upper or lower limit. The blackbody temperature, $kT_{\rm bb}$, has been estimated with {\sc diskbb} and was subsequently kept frozen during the fit.} 
\end{deluxetable*}

The best fits for Observations 2 and 3 give strong constraints on the Comptonizing medium since the electron temperature in {\sc nthcomp} is well measured. As an example, we analyze in detail the solution obtained for Observation 2. {\sc nthcomp} uses the slope, $\Gamma$, as the second parameter but for a given electron temperature, the optical depth is uniquely defined for a fixed geometry. Using the formulae given in \citet{1991MNRAS.249..634C} for a point-like source of photons and a spherical geometry of the Comptonizing medium (the same as adopted in {\sc nthcomp}), we obtain the optical depth $\tau = 4.0$ and the Compton parameter $y = 0.63$. The medium is therefore optically thick, but the Comptonization is still unsaturated due to the low value of the electron temperature. In the data, however, the normalization of the Compton component is much lower (below 10 \%) than 63 \% expected from $y$. This reveals that the material responsible for the fitted Comptonization does not cover the whole inner disk. In addition, the wind in this solution is strong; 84 \% of the material is removed from the disk at the ISCO. 

On the one hand, the maximum of the local disk temperature is at 10 $r_{\rm g}$ where the mass fraction removed from the disk is 40\%. On the other hand, half of the total flux is emitted at 40 $r_{\rm g}$ where the removed mass fraction is 22\%. If the hot, coronal material is predominantly inflowing (as a sub-Keplerian flow), then the second value is more representative for the overall Comptonization of the disk spectrum. From this we can estimate the vertical optical depth of the inflow at a given radius using
\begin{equation}
	\Mdot_{\rm cor}(r)  = 4 \pi r H_{\rm cor} \rho_{\rm cor} v_{\rm cor} 
	~~~ \textnormal{and} ~~~ 
	\tau_{\rm cor} = \kappa_{\rm es} H_{\rm cor} \rho_{\rm cor},
\end{equation}
where we only have to assume an inflow velocity, $v_{\rm cor}$, as both the corona density, $\rho_{\rm cor}$, and the corona thickness, $H_{\rm cor}$, cancel out. Assuming a velocity of 0.1 c, one then obtains at 40 $r_{\rm g}$ an optical depth of only $\tau = 0.9$. This type of material would cover the whole inner disk. Given the low normalization of the Compton component in the data, however, it cannot be the material responsible for the fitted Comptonization. The optically thick Comptonizing medium seen in the data should rather be identified with a central hard X-ray source (see Fig.~\ref{fig:sketch}). 
In Observation 3 the optical depth, $\tau = 3.7$, of the medium is somewhat lower than in Observation 2, but otherwise the solution is very similar.

\subsection{Observational tests with GRS~1915+105}
\begin{figure}[ht!]
\centering
\includegraphics[width=0.45\textwidth]{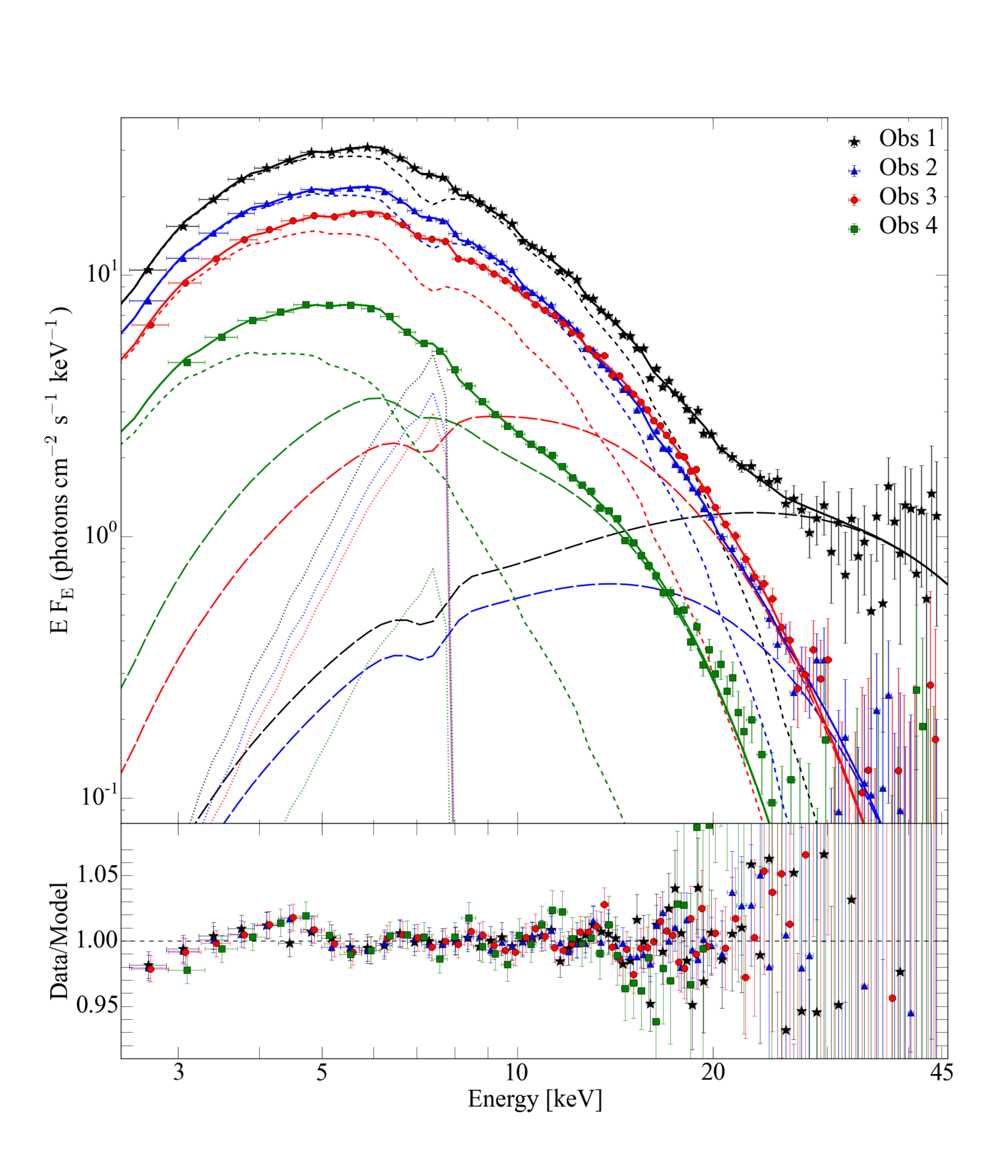}
\caption{
Four RXTE soft state spectra of GRS~1915+105 with mass accretion rates between $\approx 0.17 - 1.10 \ \MdotEdd$, fitted individually over the energy range 2.51 - 45.0 keV. The full model (solid lines) is composed of an absorbed outflow-disk component (short dashed lines), a Compton component (long dashed lines), and Fe line emission (dotted lines) and absorption features (dents at $\sim 7.2$ keV). The fit results are shown in Tab.~\ref{tab:grs1915_result}.}
\label{fig:grs1915_fit}
\end{figure}

\begin{figure}[h]
\centering
\includegraphics[width=0.45\textwidth]{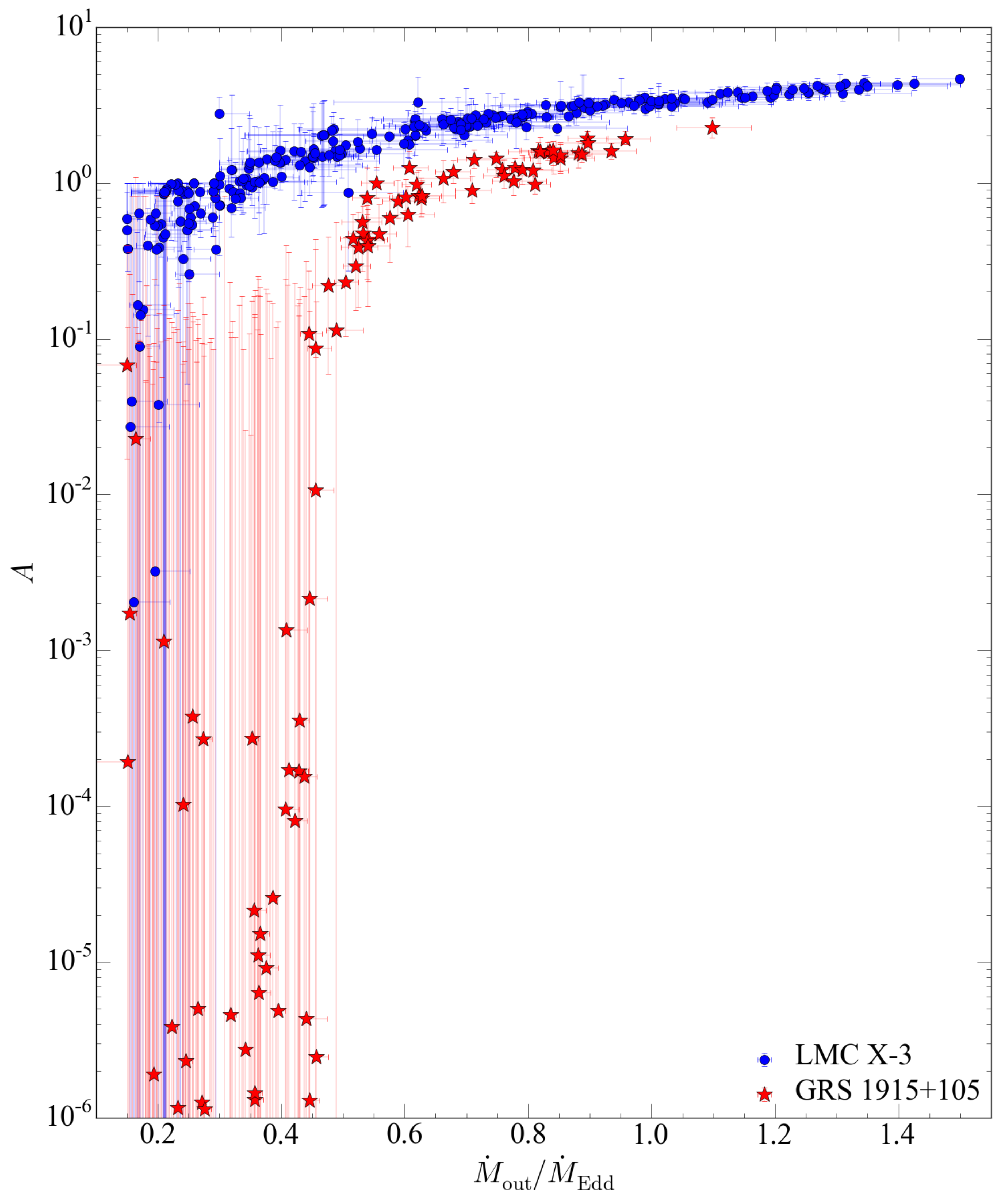}
\caption{
Required mass loss due to wind to fit a constant spin to the data. In Model~($1^*$) the quantity $A(\Mdot_{\rm out})$ parametrizes mass loss due to wind from the inner disk region. Where $A \ll 1$ no outflow is needed and the source spin is constant. }
\label{fig:amdot}
\end{figure}

RXTE made pointed observations of GRS~1915+105 from April 1996 to April 2009. We use the steady-soft observations of \citet{2015arXiv150908941P}. These energy spectra were extracted from the data of the top layer of PCU-2 and the background was subtracted using the model applicable for bright sources. A systematic error of 1\% has been added for each energy channel. From the total set of over 2000 continuous exposure segments (mean exposure time = 2.1 ks) 1257 spectra are steady and of these 264 are soft. The majority of these show absorption and/or emission as well as disk reflection features. In order to fit the soft state spectra of GRS~1915+105 we therefore include a broad Fe K emission line fixed at $E_{\rm line} = 6.4$ keV and modeled with {\sc laor} \citep{1991ApJ...376...90L} as well as a Fe absorption component ({\sc gabs}) with a central energy fitted between 6.5-7.5 keV and fixed width $\sigma = 0.5$ keV. We use {\sc smedge} to describe a smeared Fe edge where we fit the edge energy between 7 - 9 keV, leave the maximum absorption factor $\tau_{\rm max}$ free, and fix the smearing width at $W = 7$ keV. We fit the full model {\sc tbabs * (diskw + nthcomp + laor) * gabs * smedge} in the energy range 2.51-45.0 keV. 233 spectra have luminosities in our range of interest ($\dot m \geq 0.15$) and we obtain 178/233 good fits. The modeled mass accretion rates lie between $\dot m = 0.15 - 1.2$ over the range of which parameter $A$ increases from near zero to 2.5 (see Fig.~\ref{fig:amdot}). We note in particular that $A$ becomes infinitesimal already at around $45\%$ of the Eddington limit. Below, the disk is effectively a standard thin disk without winds. The electron temperature slightly increases with mass accretion rate from $kT_{\rm e} \approx 2.5 - 10$ keV together with the seed photon temperature $kT_{\rm bb} \approx 1.5 - 2.5$ keV. The latter stays roughly constant for $\dot m > 0.3$ and is substantially higher than in other Galactic XRBs but typical for the source. The absorption line energy which lies at about $E_{\rm line} \approx 7.2$ keV is practically independent of the accretion rate. Its strength, however, decreases with increasing mass accretion rate. We remind the reader that in {\sc gabs} the line strength, $N$, refers to the line depth and is related to the optical depth at the line center via $\tau_{\rm line} = N/(2 \pi \sigma)$. The anti-correlation between absorption line strength and mass accretion rate implies that an increasing amount of wind can effectively remove absorption line features. The {\sc laor} normalization which parametrizes the number of photons per area per time increases with mass accretion rate. The Fe K$_{\alpha}$ line is present in all spectra but more pronounced when wind is present. The central edge energy shows an increasing trend with mass accretion rate, $E_{\rm edge} \approx 7.25-8.75$ keV. Although the goodness of fit is already good with $\chi^2_{\nu} \sim 1$ without Fe line component, there are still small residuals which seem barely visible in some of the spectra. Adding the {\sc laor} model component to account for the Fe line is in fact improving the fits. The disk-wind model based on a fixed spin, $a = 0.95$, can fit the GRS~1915+105 data very well. We choose again four representative spectra at widely different mass accretion rates and show the data, models, and model residuals in Fig.~\ref{fig:grs1915_fit}. The results are summarized in Table~\ref{tab:grs1915_result}. 

The best fit found for Observation 1 is consistent with recent results based on data recorded with {\it NuSTAR} which operates in a wider bandpass (3 - 79 keV) than RXTE: \citet{2013ApJ...775L..45M} measure a slope $\Gamma = 2.07$ and an electron temperature $\sim$ 16 keV (see their Fig.~2) while we have only a lower limit for the electron temperature and a very simplified description of reflection due to the lack of high energy data in RXTE, which imply an optical depth of $\tau = 2.0$ for the geometry adopted in {\sc nthcomp} ($\tau = 3.1$ for Observation 1 in this work); and a similar value of the optical depth ($\tau = 1.5$) is obtained by \citet{2013ApJ...775L..45M} using the {\sc eqpair} model \citep{1999ASPC..161..375C}, where the optical depth is a free parameter. The derived Compton parameter for the {\it NuSTAR} data, $y = 0.86$ ($y = 0.91$ for Observation 1 in this work), is relatively high but, like in the case of LMC~X-3, the Compton component contributes less than 10\% to the total spectrum in the data; most of the emission stems from the disk component. This means that in Observation 1 the Comptonizing medium covers only a small part of the disk. Based on our model, the wind contribution during Observation 1 is extreme with over 90 \% of the material being removed from the inner disk. On the one hand, half of the material has left the disk by the time the disk flow reaches 6.4 $r_{\rm g}$. On the other hand, half of the flux is emitted at 14.3 $r_{\rm g}$, where the disk has lost only 26 \% of its matter to the wind. The optical depth at the latter radius is expected to be $\tau = 0.7$ for an adopted corona flow speed $v_{\rm cor} = 0.3 c$, where the high corona speed reflects the fact that the inner disk region in a high spin source like GRS~1915+105 is closer to the black hole than in a low spin source like LMC~X-3. A spherical wind would have to have a very high optical depth (see Eq.~\ref{eq:wind}), therefore, the self-consistent picture is the following: The Comptonized emission seen in Observation 1 originates in fact from a central hard X-ray source and the wind material removed from the inner disk forms an inflowing coronal layer that is accreted onto the black hole. The emission from the optically thin corona is in this scenario not directly visible in the spectral data.

The situation during Observation 4 is a quite different one. Firstly, the mass accretion rate is small and there is no significant presence of wind. Secondly, we obtain good limits on the temperature of the Comptonizing plasma since it is sufficiently low for the data to be properly recorded within the energy bandpass of RXTE. We derive the optical depth $\tau = 3.1$ from the measured electron temperature, $kT_{\rm e} = 2.35$ keV, and the lower limit of the photon index, $\Gamma = 3.25$, for the geometry adopted in {\sc nthcomp}. The corresponding Compton parameter is then only $y = 0.24$, much lower than in Observation 1, and indicates only a week Comptonization. The normalization of the Compton component is much higher, though, suggesting that a significant part of the inner disk is covered. The upper limit on the wind parameter $A$ implies an upper limit on the total wind mass loss of only 20 \% while merely 4 \% of the disk mass is liberated at the location ($r = 8.7 r_{\rm g}$) where half of the flux is dissipated. This small amount of material is not enough to form an optically thick Comptonizing zone, in particular if the material flows in as fast, as we assumed before. The coronal optical depth estimated at this radius is $\tau = 0.03$ which is much less than what is measured for the spherical Comptonizing medium. It may imply, again, the presence of a central hard X-ray source. Given the high normalization during Observation 4, however, the result rather suggests that a significant part of the Comptonization may take place in a plane-parallel optically thin corona layer. To obtain such a geometry, the hot, coronal inflow must be more than order of magnitude slower than the free fall velocity.

\begin{deluxetable*}{lcccc}
\setlength{}{0pt}
\tablecolumns{5}
\tablewidth{0pt}
\tablecaption{{\sc diskw} fits to GRS~1915+105 soft state spectra\label{tab:grs1915_result}}
\tabletypesize{\footnotesize}
\tablehead{ \colhead{model} &
\colhead{Obs 1} & \colhead{Obs 2} & \colhead{Obs 3} & \colhead{Obs 4} } \startdata
{\sc diskw} &&&& \\ [5pt]
%
$\Mdot_{\rm out}/\MdotEdd$ & 
$1.10\ ^{+0.10}_{-0.06}$ & 
$0.62\ ^{+0.05}_{-0.03}$ & 
$0.37\ ^{+0.02}_{-0.01}$ &
$0.16\ ^{+0.01}_{-0.01}$ \\ [5pt]
$A$ & 
$2.29\ ^{+0.59}_{-0.17}$ & 
$0.92\ ^{+0.25}_{-0.27}$ & 
$0.0007\ ^{+0.21}_{-0.0002}$ &
$(2 \times 10^{-6})\ ^{+0.2}_{*}$ \\[5pt]
\hline
{\sc nthcomp} &&& \\[5pt]
%
$\Gamma$ & 
$1.66\ ^{+1.12}_{-0.16}$ & 
$1.5\ ^{*}_{*}$   & 
$3.5\ ^{*}_{-0.6}$  &
$3.5\ ^{*}_{-0.25}$  \\ [5pt]
$kT_{\rm e}$ [keV] & 
$8.57\ ^{*}_{-1.81}$ & 
$3.98\ ^{+3.15}_{-0.41}$ & 
$3.94\ ^{+0.34}_{-0.71}$ &
$2.35\ ^{+0.29}_{-0.13}$ \\ [5pt]
$kT_{\rm bb}$ [keV] & 
2.08 & 
2.06 & 
2.18 &
1.77 \\ [5pt]
N [$\times10^{-2}$] & 
$0.85\ ^{+1.72}_{-0.26}$ & 
$0.50\ ^{+1.26}_{-0.15}$ & 
$4.48\ ^{+0.41}_{-1.09}$ &
$0.10\ ^{+0.02}_{-0.02}$\\ [5pt]
\hline
{\sc laor} &&& \\[5pt]
%
N & 
$0.32\ ^{+0.16}_{-0.13}$ & 
$0.23\ ^{+0.10}_{-0.09}$ & 
$0.19\ ^{+0.07}_{-0.07}$ &
$0.05\ ^{+0.01}_{-0.01}$\\ [5pt]
\hline
{\sc gabs} &&& \\ [5pt]
%
$E_{\rm line}$ [keV] & 
$7.27\ ^{+0.13}_{-0.08}$ & 
$7.24\ ^{+0.10}_{-0.07}$ & 
$7.23\ ^{+0.07}_{-0.07}$ &
$7.07\ ^{+0.09}_{-0.08}$ \\ [5pt]
N & 
$0.29\ ^{+0.09}_{-0.09}$ & 
$0.31\ ^{+0.09}_{-0.09}$ & 
$0.30\ ^{+0.08}_{-0.08}$ &
$0.35\ ^{+0.16}_{-0.11}$ \\ [5pt]
\hline
{\sc smedge} &&& \\ [5pt]
$E_{\rm edge}$ [keV] & 
$8.36\ ^{+0.39}_{-0.22}$ & 
$8.26\ ^{+0.38}_{-0.23}$ & 
$8.03\ ^{+0.30}_{-0.28}$ &
$7.37\ ^{+0.09}_{-0.09}$\\ [5pt]
$\tau_{\rm max}$ & 
$0.86\ ^{+0.28}_{-0.30}$ & 
$0.70\ ^{+0.26}_{-0.28}$ & 
$0.63\ ^{+0.20}_{-0.23}$ &
$3.06\ ^{+0.72}_{-0.88}$ \\ [5pt]
\hline \\ 
$\chi^2$/dof ($\chi^2_{\nu}$) & 
43.23/70 (0.62) & 
48.97/70 (0.7) & 
44.53/69 (0.65) &
63.47/62 (1.02) \\ [0pt] 
%
\enddata
\tablecomments{The best fitting spectral parameters. All errors are quoted at the $90 \%$ confidence level. The asterisk, $*$, indicates an unconstrained upper or lower limit. The blackbody temperature, $kT_{\rm bb}$, in {\sc nthcomp} has been estimated with {\sc diskbb} and was subsequently kept frozen during the fit.}
\end{deluxetable*}

\section{Discussion}
\label{sec:discussion}

We fitted the bulk of our thermal spectra of LMC~X-3 and GRS~1915+105 in a broad luminosity range with a simple disk-wind model that includes an inward radially decreasing accretion rate but neglects the problem of additional energy extraction from the disk by the wind. In data fitting we do not use any coupling between the amount of the removed material and the properties of the Comptonizing medium (i.e. the possible geometry, the temperature and optical depth). The model fits the data well and there is no effect of decreasing spin with luminosity since the spin is fixed by definition. The required wind is negligible at low mass accretion rates, particularly in GRS~1915+105. For increasing (outer) mass accretion rates as much as $>$ 80\% of the material can be lost in the wind (see Tab.~\ref{tab:lmcx3_result}, Tab.~\ref{tab:grs1915_result}, and Eq.~\ref{outflow1}). 

Both sources show significant Comptonization effects in the spectra. The Comptonization is consistent with the upscattering of the soft disk photons by the hot medium since the soft photon input temperature is comparable to the disk temperature. The temperature of the Comptonizing medium is higher than the maximum disk temperature by a factor 2 to 10, and in general becomes higher for higher luminosities. The power law slopes measured for GRS 1915+105 (see Tab.~\ref{tab:grs1915_result}) range from $\Gamma$ = 1.5 to 3.5 and are often not determined accurately due to the limited energy band of RXTE. The well determined slopes combined with the corresponding plasma temperatures yield the optical depths of the Comptonizing plasma of the order of $\tau$ = 2 - 3. Such a plasma phase is typical for many soft state sources \citep{1993A&A...274..105W, 1998MNRAS.301..179M, 2000Sci...287.1239Z, 2004MNRAS.353..980K, 2012MNRAS.420.1848D, 2013A&A...549A..73P}. A similar result was obtained by \citet{2013ApJ...775L..45M} with {\it NuSTAR} data, where the slope of the Comptonized component is determined more reliably than with our data. For LMC X-3 we arrive at the same conclusion: whenever $\Gamma$ is well constrained, the optical depth of the Comptonizing medium is of order of 3 - 4. The exact values depend on the geometry (plane-parallel or spherical) and the assumed photon distribution. We are using here a spherical symmetry, with photons at the center like in {\sc nthcomp}, for consistency. 

The problem is that the optical depth of the Comptonizing medium implied by the fits does not increase with the modeled disk luminosity but remains practically constant despite the sharp rise in the wind from the disk itself. Thus, the Comptonizing medium is present at low Eddington ratios, when mass loss from the disk is not required as well as at high Eddington ratios, when the implied disk mass loss is strong. Moreover, the medium does not become optically thicker as the mass accretion rate rises and winds become more prominent. This seems to suggest that the wind material is not located along our line of sight. The behavior of the normalization of the Compton component points towards an absence of Comptonizing material along the line of sight, too. Given the Compton parameter $y \sim 0.6$ (LMC X-3), for instance, the Comptonization is strong and the material should be optically thick. If this kind of material would cover large part of the disk, the disk itself would be barely visible as almost all its energy would end up in the Compton component. This contradicts the nature of the data where the Compton component makes up only a very small fraction of the total flux and the disk is seen well, including the reflected component. Therefore, the material that has been removed from the disk by the wind cannot be fully identical to the Comptonizing medium. In general, this medium likely forms a kind of magnetized dissipative skin \citep[e.g.][]{2015A&A...580A..77R}, a small fraction of it may be outflowing, but most of it would be inflowing. We sketch this geometry in Fig.~\ref{fig:sketch}. If the hard emission comes from a lamp-post style hard X-ray source, the question remains whether we have any method to prove or disprove the mass outflow from the disk. Our consistency check made for LMC~X-3 and GRS~1915+105 implies that the actual wind material may be unnoticed since the data do not require a second Comptonizing medium and the model does not overpredict the optical depth of the corona skin at the radius where half of the disk flux is emitted.

In the high spin source GRS~1915+105 we detect an absorption feature along the line of sight but we do not see it in the low spin source LMC~X-3 despite their similar inclination angles, $60^{\circ} - 70^{\circ}$. If this absorption line at $\sim$ 7.2 keV is caused by a hot plasma it must be blueshifted from its original value of 6.7 keV. This implies that the absorbing (wind) material has a speed of 0.07 c. Such a wind may originate from the innermost part of the disk but does not have to. Hydrogen-like and helium-like iron lines are resonant lines and the line-locking mechanism can accelerate the flow very efficiently independent of the launching radius. In comparison, broad absorption line (BAL) features observed in AGN have blueshifts that correspond to velocities of the order of 0.1 - 0.3 c, but the launching radius is likely further than 0.1 pc from the black hole. Recent studies of a large sample of AGN suggest typical launching radii of 10 to 1000 light days and some features can come from the distances as far as 100 - 3000 pc \citep{2013ApJ...777..168F}. The lack of line absorption in LMC X-3 might be due to its low spin. \citet{2014ApJ...793L..29S} tentatively suggested that there is a possible link between black hole spin and spectral complexity. They argued that some sources with known low spins have remarkably simple spectra with few spectral features whereas high spin sources exhibit strong Comptonisation/reflection and a rather large variety of spectral features.

Previous models of the outflows in GRS 1915+105 assumed magnetically driven outflows to reproduce the amplitude of the heartbeat regular bursts within the scenario of the radiation pressure instability \citep{2000ApJ...535..798N, 2002ApJ...576..908J}. However,  models with a static corona and a time-dependent mass exchange between the disk and the corona \citep[e.g.][]{2005MNRAS.356..205J} or models with different viscosity prescription \citep[e.g.][]{2006MNRAS.372..728M} worked equally well. 

Model~(4) which we used in Sec.~\ref{sec:flow} as one of the simple parametric prescription implicitly refers to a magnetic outflow. In this model, formally, all the matter is outflowing at the disk truncation radius. From an observational point of view, by fitting the broad asymmetrical Fe K$_{\alpha}$ line, \citet{2012ApJ...752L..21C} find that the accretion disk in 3C 120 might be truncated at $r_{\rm in} = 11.7 r_{\rm g}$ independent of the spin configuration and at a moderately high luminosity $L = 0.23 \LEdd$. The authors suggest that the material in the inner disk, instead of transforming into an advection dominated accretion flow (ADAF), could be ejected in form of a jet. This could explain the observed periodic dips in X-ray luminosity that are accompanied by large radio bursts \citep{2011ApJ...729...19K}; From a theoretical point of view, models with an inner truncation radius are frequently considered, but in the context of a two-zone geometry where an inner hot optically thin and geometrically thick ADAF is enclosed by a truncated outer standard disk \citep{1997ApJ...489..865E,2005ApJ...620..905Y,2009ApJ...707..233L,2013ApJS..207...17L,2014ARA&A..52..529Y}. In this case, the ADAF fully describes the low luminosity/hard emission state where the accretion rate drops below a few percent of the Eddington limit \citep{2004MNRAS.351..791Z}. However, this trend goes in the opposite direction compared to the trend required to solve the spin paradox in our paper: the transition radius in the case of standard disk/ADAF hybrid solution decreases with the increase of the Eddington ratio while we need an opposite trend. Therefore, the mechanism that acts in high Eddington ratio sources must be a different one. Observationally, the presence of the relativistically broadened reflection provides the support for the disk close to the black hole. We see it in some of the RXTE spectra of GRS~1915+105, and soft state reflection has been detected in some other galactic sources in the soft state, e.g. Cyg~X-1 \citep[][]{2014ApJ...780...78T}; XTE~J1908+094 \citep[][]{2015ApJ...811...51T}; LMC~X-1 \citep[][]{2015PASJ...67...46K};  4U~1543-47 \citep[][]{2014ApJ...793L..33M}, as well as in AGNs, e.g. Mkn~335 \citep{2015MNRAS.454.4440W}. In this case only a small fraction of material should be outflowing, but this is not fully self-consistently treated in the Model (4) as the temperature of the disk with magnetically driven wind should be much lower than the one for the absence of the $f_{\rm wind}$ term in Model (4), which affects the color-correction term. Therefore, we do not use this model when fitting the observational data in 
Sec.~\ref{sec:tests}.

\subsection{Comparison between disks with and without wind}
There is only one significant difference between disks with and without wind, namely the predicted mass accretion rate. For a given observation, a disk with initial mass accretion rate modified by wind can exhibit accretion rates significantly larger than a disk without wind. This is a direct consequence of $\dot m$ being measured at the outer disk edge and not being constant with radius due to wind mass loss. Accretion disks that self-consistently include winds are thus much less efficient at converting mass into radiation than a standard thin disk.

The spectral difference in the shape between the models with and without wind is very small, when the model parameters are appropriately adjusted. In a given observation, the spin, accretion rate, and wind prescription are strongly degenerate. They are additionally masked by the Comptonization at higher energies. A sequence of disk-dominated spectra is necessary, since then the spin should remain the same, independently from the source luminosity. Models without wind and with hardening factor taken from {\sc bhspec} do not provide the correct solution for such a sequence for GRS~1915+105 and LMC~X-3, while the wind model presented in this paper is satisfactory.

The comparison of the disk wind model with the data is a first step towards the assessment of the flow geometry. However, next steps are very difficult, since the material removed from the disk flows out as a wind but also (mostly) flows in as a corona flow. Comptonization computations should be done in 3-D, for the adopted flow pattern. The basic difficulty is that we then need to know the temperature distribution inside the hot flow which depends on the plausible local dissipation, and the global magnetic field will affect the global flow.

The success of the wind model in reproducing the right trends over a range of mass accretion rates does not mean that the wind is the only solution to the spin problem. \cite{2011A&A...533A..67S} have shown that if a constant hardening factor is used instead of a variable hardening factor predicted by {\sc bhspec} and {\sc slimbh}, the spin problem is significantly alleviated. We find that fits with {\sc slimbh} are just as good as those with {\sc diskw} and, consequently, that the RXTE data are not sensitive to the shape of the disk model spectrum. {\it Suzaku} \citep{2007PASJ...59S...1M} is a satellite that has a moderate spectral resolution CCD for the 0.7-10 keV band and a hard X-ray detector for the 12-30 keV band. \citet{2010ApJ...714..860K} used {\it Suzaku} observations of LMC~X-3 to constrain the spectral shape of the intrinsic disk emission in various models. They found that {\sc bhspec} which self-consistently includes radiative transfer through the vertical structure of the disk gives an excess at the absorption edge energy below 1 keV compared to the data. This may suggest the need to include more physical processes in the disk atmosphere, i.e. atomic physics and better assumptions for the disk density and emissivity profile with respect to the disk height.

One way to prove the physical existence of the wind close to the black hole horizon would be to see it directly in the spectral shape of the pure disk component, due to general relativity effects. However, this would require a data set with very low systematic errors, below 0.2\%. In addition, the precision in the description of the Comptonization process would also matter in this case, including the 3-D distribution of the static Comptonizing medium and taking into account the material motion would pose yet another challenge \citep{1999ApJ...510L.123B}.

\subsection{Observational evidence of winds}
Outflows are expected and actually observed in accreting systems. Their occurrence in the form of collimated jets is characteristic of low Eddington rate sources. While jets seem to be suppressed in the high/soft state of black hole XRBs \citep{2015MNRAS.448.1099Q}, uncollimated outflows/winds are still found in high luminosity sources \citep{2006ApJ...646..394M, 2014ApJ...784L...2K}. However, the presence of the specific outflow we request here to solve the problem of the apparent spin decrease is difficult to test. The outflow temperature must be at least $10^7$ K or higher and the outflow may, or may not, be located along the line of sight, depending on the collimation which is unspecified. The material is thus highly, or fully ionized leaving only a possibility of emission/absorption from highly ionized iron. 

There are ample observational evidences for outflows from accretion disks. Outflowing winds have recently been detected in several XRBs via blueshifted X-ray absorption lines which are observed in high resolution spectra \citep[see][and references therein]{2015ApJ...814...87M}. With modeling the observed narrow absorption features recorded in GRS~1915+105 by Chandra HETGS (High Energy Transmission Grating spectra), the total mass outflow rate is estimated to be comparable to the mass accretion rate in the inner part of the disk \citep{2009ApJ...695..888U}. A re-analysis of the Chandra HETGS spectroscopy data of GRS~1915+105 by \cite{2015ApJ...814...87M} with improved multi-zone photoionization models and with different ionization parameters and velocities reveals that $\Mdot_{\rm wind}/\MdotEdd \simeq 0.3$ when the accreted gas rate $\Mdot_{\rm accr} \simeq 6 \MdotEdd$, where the Eddington accretion rate in that work is defined as: $\MdotEdd \equiv \LEdd/c^2$. 

These detected winds, however, clearly do not correspond to the winds requested in our paper to solve the problem of the apparent spin evolution. Our winds should happen very close to the black hole, at a few gravitational radii, and with the terminal speed of the order of the local Keplerian velocity at these radii, i.e. at a large fraction of the light speed. All winds discussed above show narrow absorption features with shifts corresponding to a few hundreds to a few thousands km s$^{-1}$, e.g., 4U~1630$-$472: the velocity $v < 8.0 \times 10^{-3} c$ and the launching radius $r \sim 800 - 6100 r_{\rm g}$; GRO~J1655$-$40: $v < 11.8\times 10^{-3} c$ and $r \sim 500 - 10000 r_{\rm g}$; H~1743$-$322:  $v < 4.3 \times 10^{-3} c$ and $r \sim 1100 - 4900 r_{\rm g}$; GRS~1915+105: $v < 4.0 \times 10^{-3} c$ and $r \sim 1200 - 23000 r_{\rm g}$ \citep[see Table 4-6 and 12 of][]{2015ApJ...814...87M}. Thus, observations usually give constraints on the partially ionized winds from the outer disk while here we deal with the highly ionized winds from the inner disk. 

Magnetically driven outflows have been postulated and studied from a theoretical point of view \citep[e.g.][]{1982MNRAS.199..883B,1994ApJ...434..446K,1994ApJ...429..139C}, and efficient outflow close to the inner disk radius due to the bending of the poloidal field is expected \citep{2010MNRAS.401..177C}. Observationally, magnetic assistance has been advocated in some outflows, like ultra-fast outflows in active galaxies. Ultra-fast outflows have been detected in a number of AGN through the analysis of absorptions features. For example, in PG 1211+143 \citet{2003MNRAS.345..705P} identified highly ionized outflows which had a velocity of $\sim 0.08 - 0.1 c$ and mass flux comparable to the mass accretion rate. A more recent spectral analysis of XMM-Newton/EPIC data of PG~1211+143 revealed an outflow velocity of $\sim 0.1 - 0.2 c$, with most of the wind launched at 200 $r_{\rm ISCO}$ and an inner wind truncation radius of $30r_{\rm ISCO}$ \citep{2015ApJ...805...17F}, which favors magnetic driving. The location of this wind, however, is still too far from the black hole for our purpose. 

In principle, the motion of the plasma is encoded in the way how a fully ionized plasma up-scatters the photons. Since the terminal outflow velocities are large while the initial denser part of the wind has a lower velocity, the expected effects of the bulk Comptonization are difficult to model, without assuming a specific velocity profile. The issues of bulk Comptonization were occasionally discussed both in the case of an inflow (e.g. Chakrabarty \& Titarchuk 1995; Liu et al. 2015) and an outflow \citep[e.g.][]{1999ApJ...510L.123B}. Computations of the Comptonization of the disk photons in the postulated winds with density/velocity/temperature gradients is beyond the scope of the present simple work.

\section{Conclusions}
\label{sec:conl}
The determination of the black hole spin in XRBs using advanced fully relativistic disk models including full radiative transfer to account for Comptonization in the disk atmosphere leads to a paradox: the spin decreases with increasing source luminosity \citep{2006ApJ...652..518M, 2010ApJ...718L.117S, 2011A&A...533A..67S}. Here, we analyze whether this effect might be explained by the presence of the winds from the innermost part of the accretion disk. We find that relatively smooth radial winds without explicit dependence on mass accretion rate (Models (1)-(3) in Sec.~\ref{sec:flow}) do not provide the solution for this effect. However, if we assume in Model $(1^*)$ that parameter $A$ depends on $\Mdot_{\rm out}$ (the mass accretion rate at the outer disk radius) the apparent spin decrease can be reproduced (green stars in Fig.~\ref{fig:1915}). We can recover the observed trend also in Model~(4) where the truncated inner disk radius depends linearly on $\Mdot_{\rm out}$. The physical interpretation of such a model is that either most of the material, or at least most of the energy and angular momentum, has to be removed from the disk in a form of a wind or an uncollimated jet, leaving behind a cold non-radiating flow close to the black hole horizon. 

In all these scenarios the shape of the thermal disk component is modified, leaving a trace of the process in the spectral shape. We show that the spectra for the two cases of outflow discussed in Sec.~\ref{sec:comp}, namely the hot optically thin and the cold optically thick wind, are in comparison to a spectrum without Comptonization either visibly harder or softer, respectively. However, if the Comptonizing medium is chosen independently from the predicted outflow, the difference between the model with, and without outflow is very small, $\sim 0.2\%$ at the energy where the disk flux before Comptonization drops by a factor of two below the peak flux (see Fig.~\ref{fig:spectra2}). 

We test in more detail the disk-wind scenario against the data for two bright XRBs, LMC~X-3 and GRS~1915+105, using the Model $(1^*)$ and assuming a Comptonizing medium surrounding the disk. For each source we fit several hundreds of high/soft spectra assuming an intrinsically constant black hole spin and the presence of the wind that correlates with mass accretion rate. We obtain good fit statistics as demonstrated for four representative spectra for each source (Figures~\ref{fig:lmcx3_fit}, \ref{fig:grs1915_fit} and Tables~\ref{tab:lmcx3_result}, \ref{tab:grs1915_result}). The apparent decrease of the black hole spin over a range of mass accretion rates in LMC~X-3 and GRS~1915+105 could originate from the presence of wind or be a consequence of an incomplete description of the disk atmosphere. The latter issue will be addressed in a forthcoming paper. Here, we show that the wind model with a fixed spin can fit the data. 

However, the implied properties of the Comptonizing medium do not correlate with the required wind, as the wind intensity rises sharply with the luminosity while the Comptonizing medium properties remain almost unchanged. This implies the presence of a separate Comptonizing medium, likely in the form of an X-ray source located on the symmetry axis, while the material removed from the disk, albeit in large amount, can escape detection if it flows predominantly inward as a corona flow. Thus we have no direct observational evidence of the requested wind from the inner few gravitational radii.

\section*{Acknowledgments}
Our warmest and most heartfelt thanks go to R. Remillard and J. F. Steiner who shared their data for this study. We thank F. Yuan for his helpful comments. Furthermore, we are greatly indebted to the anonymous referee whose to the point comments and suggestions helped to significantly improve the manuscript. The project has received funding from the European Union Seventh Framework Program (FP7/2007-2013) under grant agreement No.~312789. This work was also supported by the Ministry of Science and Higher Education grant W30/7.PR/2013 and the Polish National Science Center grants No.~2011/03/B/ST9/03281 and No.~2013/08/A/ST9/00795.





\end{document}